# Production-ready double-side fabrication of dual-band infrared meta-optics using deep-UV lithography

*Kai Sun\*, Xingzhao Yan, Jordan Scott, Jun-Yu Ou, James N. Monks, Otto L. Muskens\**

Kai Sun, Jordan Scott, Jun-Yu Ou, Otto L. Muskens
Physics and Astronomy, Faculty of Engineering and Physical Sciences, University of Southampton, Southampton, SO17 1BJ, United Kingdom
E-mail: K.Sun@soton.ac.uk, O.Muskens@soton.ac.uk

Xingzhao Yan
Optoelectronics Research Centre, Faculty of Engineering and Physical Sciences, University of Southampton, Southampton, SO17 1BJ, United Kingdom

James N. Monks
Teledyne Qioptiq ltd, Glascoed Rd, Saint Asaph, LL17 0LL, United Kingdom

Meta-optics, the application of metasurfaces into optical systems, is seeing an accelerating development owing to advantages in size, weight and cost and the ability to program optical functions beyond traditional refractive optics. The transition of meta-optics from the laboratory into applications is enabled by scalable production methods based on highly reproducible semiconductor process technology. Here, we introduce a novel method for fabrication of double-sided metasurfaces through deep-UV lithography as a production-ready method for achieving high-quality meta-optics. We achieve patterning of a silicon wafer on both sides with mutual alignment of around 25 µm based on tool accuracy, without requiring through-wafer alignment markers other than the wafer notch. A first novel application highlighting the benefits of double-sided design is demonstrated in the form of a dual-band metalens with independent control over focal lengths in mid- and long-wave infrared bands. Using multi-reticle stitching we demonstrate a 40 mm diameter, large-area metalens with excellent broadband imaging performance, showing partial cancelling of chromatic dispersion when used in a hybrid configuration with a BaF$_2$ refractive lens. Our work opens new avenues for infrared meta-optics designs and double-side meta-optics fabrication through a production-ready technique which can be directly translated into scalable technology for real-world applications.

1. Introduction

Metasurfaces consisting of sub-wavelength nanostructures allow for engineering of optical responses beyond that of the bulk material properties including amplitude, phase, polarization and absorption.[1-4] Metasurface optics, or in short meta-optics, have gained significant attention with research progress in recent years, and are commonly referred to as 'flat optics' for the thickness being essentially that of the nanostructured layer and its substrate, which remains unchanged with increasing diameter. Meta-optics including metalenses could be highly desirable over bulk refractive solutions in terms of size and weight, materials selection, thermal, and mechanical stability,[5] which are critical parameters in applications such as space and low-form factor platforms such as smartphones and wearables. In the past decade, meta-optics have been intensively investigated in the visible range[6-10] and near-infrared and short-wave infrared (SWIR) bands[11-15] for potential applications in a wide range of areas including AR/VR, 3D imaging, facial scanning, lidar, and astronomical telescopes, to name a few. Metalenses in this spectral range are mainly formed through layers of amorphous silicon, gallium nitride, or titanium dioxide nanostructures on transparent substrates e.g. glass/quartz and sapphire, with the nanostructure fabrication as the main challenge.

While most meta-optics laboratory demonstrations were made by using electron-beam lithography[13, 16] and two-photon laser writing,[17-18] increasingly progress has been made on the use of scaled-up fabrication techniques such as deep-UV lithography[15, 19-20] and nanoimprint lithography.[21-22] The latest developments show promising trends, such as the increase in metalens diameters from a few ten micrometers[23] to 100 mm through multi-exposure stitching technique,[24] whilst the substrate size is also increased from a few mm chip to 8-inch and even 12-inch wafers in 2024.[19-20] New functions such as multi-band and multi-wavelength operation were also reported to further increase the metalens capabilities.[7, 25-27] Double-sided metalenses were investigated with one metasurface on each side of the substrates for their improved capability and performance.[6, 10, 28-30] Unlike the single-sided metalenses, double-side metalenses involve further challenges in ensuring defined nanostructures are preserved through the subsequent lithography steps. Currently, the demonstrated works only include small-diameter visible and SWIR metalenses fabricated through e-beam lithography, as the DUV lithography offers little flexibility in handling nanostructures as the backside and also has strict substrate requirements e.g. bowing and thickness variation, to achieve the specified minimal feature controls.

Compared with the decade-long investigation on visible and SWIR metalenses, research focusing on long-wave infrared (LWIR) metalens solutions has been reported much more recently.[31-38] These demonstrations are done mainly using all-silicon implementations, owing to the high refractive index and low absorption in LWIR bands, and matured semiconductor processing methods using e-beam lithography, laser writing, and DUV lithography. At long wavelengths the nanostructures or pillars are a few micrometers in height and thus DUV lithography might be the most suitable technique over nanoimprint for scale-up manufacture. Large area LWIR metalenses =up to 40 mm have only recently been reported through direct laser writing[33], while mid-wave infrared (MWIR) metalenses so far have received limited attention.[39-41] An overview of the recent literature with parameters relevant to this work is presented in Table S1, Supporting Information.

In this work, we demonstrate a novel wafer-scale double-side metasurface fabrication method on 200 mm silicon substrates, offering full compatibility with DUV lithography and automated wafer coater and developer, commonly referred to as the Track. Based on our novel double-sided fabrication method, we have developed novel MWIR/LWIR dual-band metalenses with operation wavelengths of 4.0 µm and 10.0 µm with independent control of the focal length at each wavelength. We further demonstrate that the double-side process is compatible with size scale-up through a multi-exposure stitching technique and we have successfully achieved a metalens with diameter of 40 mm. This technique has been applied to present large-diameter MWIR/LWIR dual-band and LWIR single-band metalenses and show their real-world performance.

Large-diameter infrared metalenses manufactured through scale-up production technologies are urgently needed to advance the emerging infrared imaging sector. Infrared imaging is gaining significant popularity for its increasingly broad application in sectors covering autonomous driving, road safety, industrial monitoring, gas sensing, medical imaging, building energy efficiency, wildlife monitoring, satellite earth observation, defence, security, and space applications. The LWIR and MWIR bands are two critical atmospheric transmission windows, which respectively correspond to the near-room-temperature blackbody emission peak for LWIR and the higher temperature emission for MWIR, as well as absorption wavelengths of several gases e.g. methane ($CH_4$), nitric oxide ($N_2O$), carbon dioxide ($CO_2$). For LWIR optics, most classical optical materials such as oxide glass and polymers are opaque due to phonon absorption, leaving the conventional solutions to single-crystal germanium and chalcogenide glasses like zinc sulfide (ZnS) and zinc selenide (ZnSe). These materials are either high-cost and hard to source (Ge) or have limited mechanical robustness (ZnS and ZnSe). For MWIR optics, there are similar material challenges. Moreover, Ge lenses suffer from the thermo-optical focal drift and ZnS and ZnSe lenses have limited thermal shock resistance and are fragile against scratching. Meta-optics could offer a cost-effective alternative able to address some of the open challenges in infrared imaging technology. Our work in this study offers a number of promising opportunities for double-sided infrared meta-optics.

2. Results

1.2 Fabrication of double-sided meta-optics using deep-UV lithography

To achieve the production-ready fabrication of double-sided meta-optics, the key challenge is to ensure its compatibility with automatic handling through DUV lithography and Track systems. The proposed method is schematically shown in Figure 1A, with the key element being the use of a double-hard masking method. The process flow starts with a sub-1 µm $SiO_2$ layer growth on both sides of double-side polished (DSP) silicon substrates through thermal oxidation (step 1). The DUV lithography was subsequently performed on one substrate side referred to as front side (step 2). After the lithography, the $SiO_2$ layer is plasma etched as the hard mask for subsequent Si etch (step 3). After the resist removal, the wafer is flipped over with patterned Front side down and the second DUV lithography is performed on the other substrate size referred as back side (step 4).

Here, the $SiO_2$ layer thickness is crucially chosen to be below 1 µm, which makes patterned $SiO_2$ structures consistent with the unpolished side of standard silicon wafers with a typical roughness below 1 µm. The DUV lithography and Track systems are capable of handling the patterned $SiO_2$ side directly without needing any modification. The front-to-back side alignment is handled by the DUV lithography system through substrate notch identification. The back side $SiO_2$ is then also patterned using the same plasma etching (step 5). After this step, the most critical lithography steps have been completed with the hard mask $SiO_2$ layers on both sides well-patterned. Subsequently, the patterns are transferred from the $SiO_2$ hard mask into the Si substrate through plasma etching in two steps, one for back side (step 6) and one for front side (step 7). After the Si etching, the SiO$_2$ hard mask on both sides is removed through a wet etch in buffered hydrogen fluoride (HF), leaving double-side metasurface over Si substrates (step 8). The proposed method is capable of manufacturing large-diameter metalenses beyond the exposure size limit (33 mm × 25 mm) of the DUV lithography tool used (Nikon NSR-204B scanner, 248 nm DUV wavelength) through a multi-exposure stitching step over pre-defined alignment marks (Section S2, Supporting Information).

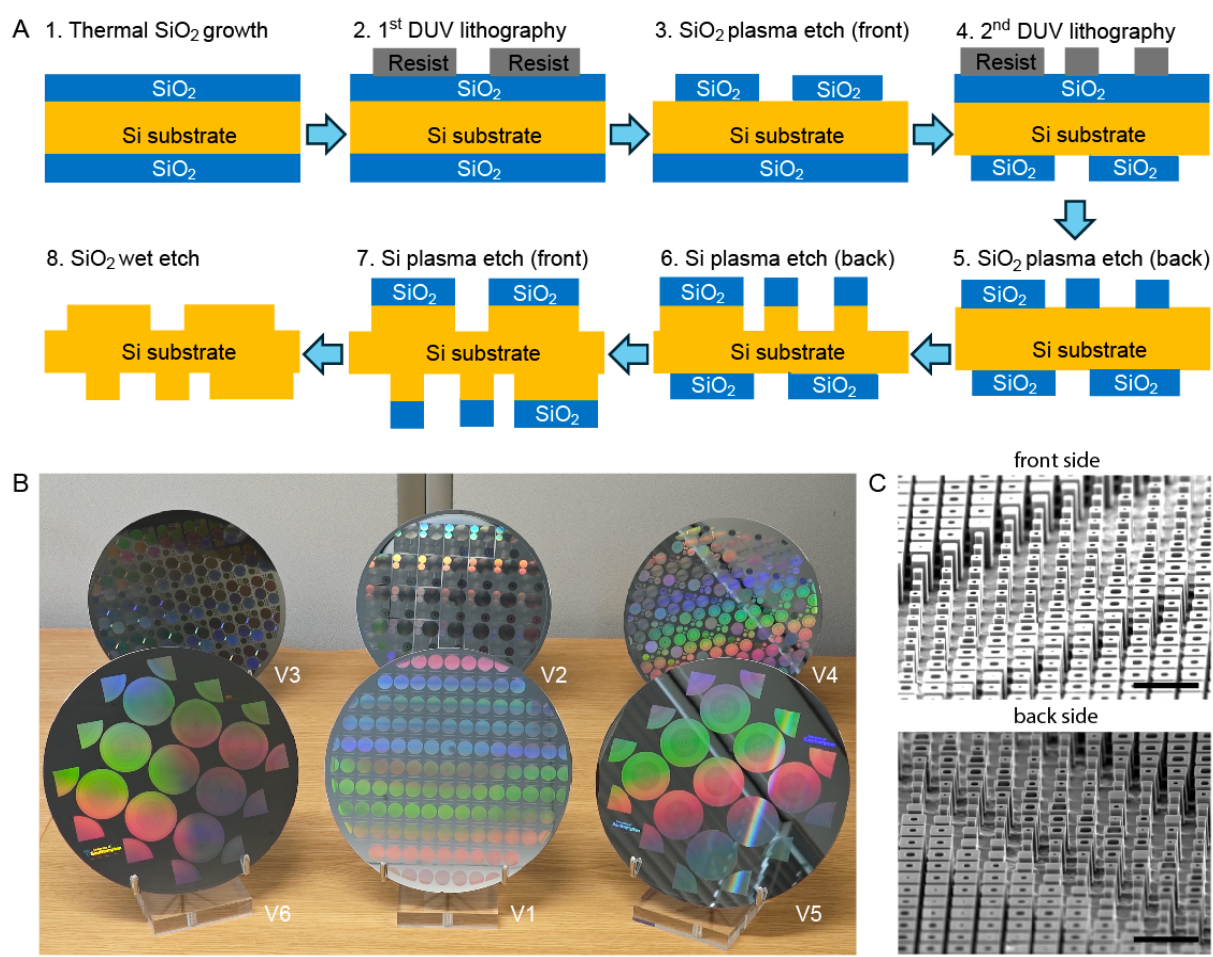

**Figure 1. Wafer-scale manufacture of double-side meta-optics through deep-UV.** (a) Process workflow for fabrication of double-sided meta-optics using deep-UV lithography and plasma etching. (b) Family photograph of six generations of meta-optics fabricated on 200 mm silicon wafers labelled V1-6. (c) SEM images taken at 65° sample tilt from representative arrays at front and back of wafer V5. Scale bars, 5 µm.

Figure 1B shows a photograph of the first six fabricated double-sided substrates by our team with meta-optics of different diameters and functions. The fabricated metalenses have diameters ranging from a few millimeters up to 40 mm. Results from wafers V4 – V6 in particular, will be discussed in this work. Scanning electron microscopy (SEM) images of the front and back sides after processing are shown in Figure 1C. In this particular design, hollow pillars were used with minimum feature sizes of around 200 nm given by the 248 nm DUV fabrication process. The etch depth on both sides was nominally 3.5 µm, with some variation dependent on gap size due to microloading effect (Section S3, Supporting Information). This etch depth meets MWIR band requirements even for single-side meta-optics, while the double-side meta-optics geometry also ensures that total etch depths of 7.0 µm match the requirement for phase control in the LWIR band.

In this method, the accuracy of registration of features on both sides of the substrate is defined by the mechanical accuracy of the tool, with the only through-wafer marker being the wafer notch provided by the substrate manufacturer. Accuracy and reproducibility of the wafer notch, in combination with notch recognition features on the DUV tool, are therefore critical for obtaining the alignment of the double-sided meta-optics components. Alignment measurement patterns were included on the DUV reticles which allow verification of registration after the fabrication using an infrared camera. Results (Section S4, Supporting Information) indicate that alignment of around 25 µm between front and back side is achieved, with the main cause for misalignment being a wafer rotation of around 0.045°. either due to an asymmetry in the wafer notch or a tool calibration limit. This rotation results in a misalignment that is increasing away from the center of the wafer, and best results were obtained for lenses in the center of the wafer. Subsequent tests indicate that it is indeed possible to increase or reduce the rotational misalignment within a given batch of wafers by applying an extra wafer rotation to the second exposure. Future work will be needed to assess the reproducibility of systematic corrections over many production batches. A 25 µm misalignment of front to back side amounts to 3.4% of the 725 µm substrate thickness, which translates into a skewness of the double-sided meta-optic with respect to the optical axis of 2°.

2.2 Design of dual-band meta-optics

To demonstrate the functionality of our double-side meta-optics platform, we target the design and fabrication of a dual-band metalens operating simultaneously at MWIR and LWIR bands. This functionality is achieved by a joint optimization of the two metasurfaces, which is computationally implemented using a commercial simulation platform (RSoft Metaoptics Designer, Synopsys) using a bidirectional scattering distribution function (BSDF) database generated by using the rigorous coupled wave analysis (RCWA). Two design wavelengths were selected as 4 µm for MWIR and 10 µm for LWIR. Unit cell periods between 1.8 – 2.6 µm were tested and a period of 2.2 µm was found to offer a good compromise of performance at both wavelengths.

As a basic design concept we use two independent length scales in a single unit cell to control the two different wavebands. Symmetric structures such as hollow squares and crosses allow polarization independent performance of the metalens. Compared to crosses, hollow squares furthermore provide two short-axis resonators within each unit cell reducing the diffractive losses in the MWIR band. A crossover from hollow squares to filled squares and crosses was defined in a simple parametrization using four rectangular segments of length L and width W placed at offset

$$\Delta_{x,y} = \frac{W}{2} - \frac{H}{2}, \quad (1)$$

where $\Delta_{x,y}$ denotes the offset in either x or y as shown in **Figure 2**A. We vary L from 0.2 – 1.8 µm and W from 0.2 – 0.9 µm with fixed pillar height of 3.5 µm, resulting in the maps shown in Figure 2B at 4 µm and 10 µm wavelength for transmission and phase for period metasurface arrays on a single silicon-air interface. The phase map at 4 µm spans multiple orders of $2\pi$, showing a large variation in available design conditions even when using a single metasurface layer. While such a large phase diversity is needed to achieve independent control over the optical phase within the two bands, strong variations will increase sensitivity to fabrication tolerances as well as resulting in increased wavelength dependence of performance. We also observe that there are considerable regions of parameter space where transmission is low at 4 µm wavelength, in particular for the larger filled squares due to a breakdown of the design principles requiring at least one small length scale compared to the wavelength. At 10 µm wavelength, transmission is overall high ranging between 85% - 96%, while the optical phase covers roughly half of the required full $2\pi$ range, thus necessitating the use of two metasurface sides to achieve its full optical functionality.

As the metasurface optimization prioritizes lens transmission against desired point spread function at both wavelengths, it should converge towards regions with higher transmission within a certain penalty range for losing the ideal phase response. An example of this is shown in Figure 2D, which plots the probability P for each element in the database to be selected in a 40 mm diameter dual-band lens with $3.2\times10^8$ elements. Results are shown for the front side, with very similar results obtained for the back side (Figure S9, Section S5, Supporting Information). Overall, the optimization algorithm selected predominantly hollow squares (56.6%), followed by filled squares (41.1%) and only very few crosses (2.3%). Areas with higher probability and those which are mostly avoided can be located in the parameter map. Larger rings located on the right-hand side of the parameter space are often required in the design to access the higher optical phase shifts for LWIR, at the cost of some MWIR transmission.

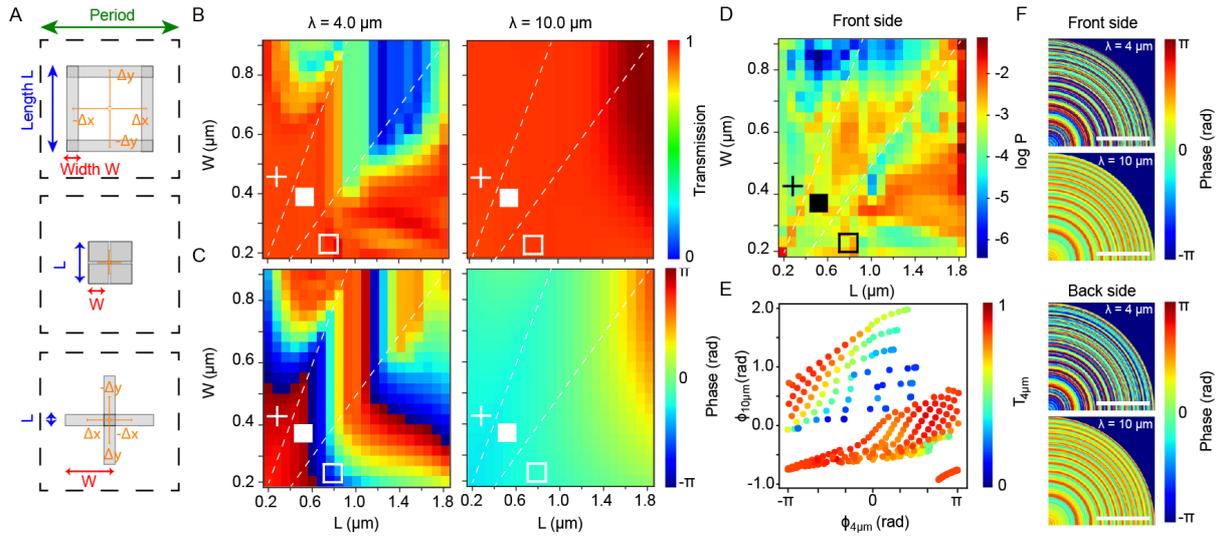

**Figure 2. Design of double-side dual-band metalenses operating at λ = 4 and 10 μm.** (a) Definition of design space with period *P*, length *L* and width *W* and transition from hollow and filled squares to cross shapes following Eq. 1. (b,c) Design space showing transmission and phase response at two wavelengths λ of 4 μm (b) and 10 μm. Dashed lines indicate transitions between shapes. (c) Probability (log-scale) for selection of each design element in the large-area dual-band metalens (V5). (d). (d,e) Parametric plot of phase $\phi_{4\mu m}$ and $\phi_{10\mu m}$ for the design space. (f) Example phase profiles (quadrant) for small-diameter (2 mm) F/2 lens showing generated front and back side profiles at λ = 4 μm and 10 μm. Scale bars are 0.5 mm.

The diversity of elements selected in the optimization process is indicative that the dual-band design requires the substantial design space in order to satisfy simultaneously phase combinations at both wavelengths. A different way of presentation this space is through a parametric scatter plot of the optical phases $\phi_{4\mu m}$ and $\phi_{10\mu m}$ for each element, as shown in Figure 2E. Here we see that for a single metasurface, indeed a wide range of combinations for $\phi_{4\mu m}$ and $\phi_{10\mu m}$ can be addressed. Each data point is furthermore color coded according to their transmission at 4 μm wavelength, to illustrate the regions of low MWIR transmission (blue data points). Compared to the single metasurface, double-sided meta-optics allow different combinations of phases to be constituted by choosing appropriate linear combinations of phase elements for each side, taking into account the wave propagation inside the substrate, thus offering a much wider range of choices for the dual-band design than for a single-side meta-optic device. A typical example of a double-sided optimized design is shown in Figure 2F showing the resulting optical phase profiles at 4 µm and 10 µm wavelength, for a small 2-mm diameter lens. The small diameter was chosen to allow resolving the individual rings, however, results are very similar for much larger lens diameters. The phase profile at 10 µm resembles that of a traditional lens with a hyperbolic phase profile and approximately half the required phase on each side, while the profile at 4 µm shows much more rings with varying phases, that cannot be easily traced back to a particular functional shape.

**2.3 Fabrication of dual-band metalenses with independent focal length control**

Using the metasurface database of Section 2.2, we proceed with developing a set of dual-band metalenses with diameters of 12.5 mm. Three different designs are compared, where we select a fixed focal length of 25 mm (f-number F/2) at 4 µm wavelength and a focal length of 20 mm, 25 mm and 30 mm at 10 µm wavelength, resulting in f-numbers of F/1.6, F/2.0 and F/2.4, respectively. **Figure 3**A illustrates the design concept with the three LWIR focus conditions color coded in blue, green and red. The different lens designs are all accommodated on a single 25 × 33 mm$^2$ reticle to allow the fabrication on the same wafer, thus eliminating any variation in fabrication parameters between designs. Figure 3B shows the front and back sides of the fabricated wafer, containing >100 individual metalenses.

Focal lengths in the MWIR range were determined using a 1951 USAF positive test target on glass in a reverse configuration, where the meta-optic was used as an imaging objective and the achromatic camera lens as a tube lens. As the illumination source, light from a silicon nitride membrane blackbody source at a temperature of 450°C was spectrally filtered using a set of bandpass filters with center wavelengths of 3.25, 3.75, 4.25 and 4.75 µm and bandwidth of each filter of 0.5 µm. Resulting images at λ=4.25 µm of group number 3 (elements 1 and 2) are shown in Figure 3C. A high-frequency enhancing filter is applied to all images in this study as a very simple haze correction requiring no prior knowledge of the metalens (Section S6, Supporting Information). The filter suppresses lower spatial frequency components below 3 lines per mm in the MTF as shown in Figure 3D, reducing contributions from zero-order transmission and chromatic-aberration induced blurring. More advanced dehazing algorithms may be implemented beyond the scope of this demonstration[42].

Focal lengths at LWIR were determined by direct imaging of the surface of the 2.1×1.8 mm² SiN light source itself, as the USAF substrate did not transmit sufficient light to enable its use at longer wavelengths. Bandpass filters were used with center wavelengths of 9, 10, 11 and 12 µm and bandwidth of 0.5 µm, results for 10 µm are shown in Figure 3C. Other results were taken for the different spectral filters (Figure S16, Section S7, Supporting Information), resulting in values of the focal length $f$ against wavelength shown in Figure 3E (data points).

Clearly, the three different lenses produce different focal lengths in the LWIR and therefore different magnifications of the test image, whereas the focal length at MWIR remains unchanged. This experimental evidence confirms that the lenses successfully operate as dual-band MWIR/LWIR optical elements with independent control over the focal length and resulting in good quality images in both bands. We can directly compare these results against the model design by plotting the simulated on-axis intensity against distance $z$ as a color map for each evaluated wavelength. Here the color scale indicates the intensity normalized against the maximum in each band. Simulation results for the three different focal lengths in LWIR are well separated and could be overlayed for visualization purposes, separate results are shown in Figure S12-S14, Section S7, Supporting Information. Absolute focusing efficiencies $\eta$ within are also extracted from the simulations using a 50 µm diameter aperture located at the distance corresponding to the maximum on-axis intensity for each wavelength, and are plotted in the top panels of Figure 3E. We see that the designed lenses achieve a peak efficiency of around 35% at $\lambda$=4.0 µm with a spectral bandwidth of ±0.25 µm around this maximum. In comparison, the efficiency at LWIR is higher and reaches up to $\eta$=50% at 10 µm with a spectral band exceeding ±1 µm around the maximum. Apart from the efficiency drop, we also observe the $1/\lambda$ wavelength dependence of the focal length expected for diffractive optical elements without chromatic correction.

To measure the absolute focusing efficiency a $CO_2$ laser at 10.8 µm wavelength was used in combination with a 200 µm pinhole (Figure S17, Section S7, Supporting Information). All three lenses show an absolute transmission of 46% and relative focus efficiency of between 50%-55%, yielding absolute focus efficiencies of 23%-26%. Numerical design simulations in Figure 3E indicate an ideal efficiency of around 30%, which remains largely the same for collection apertures above 50 µm. Therefore, our lenses perform within the expected range of efficiencies at LWIR.

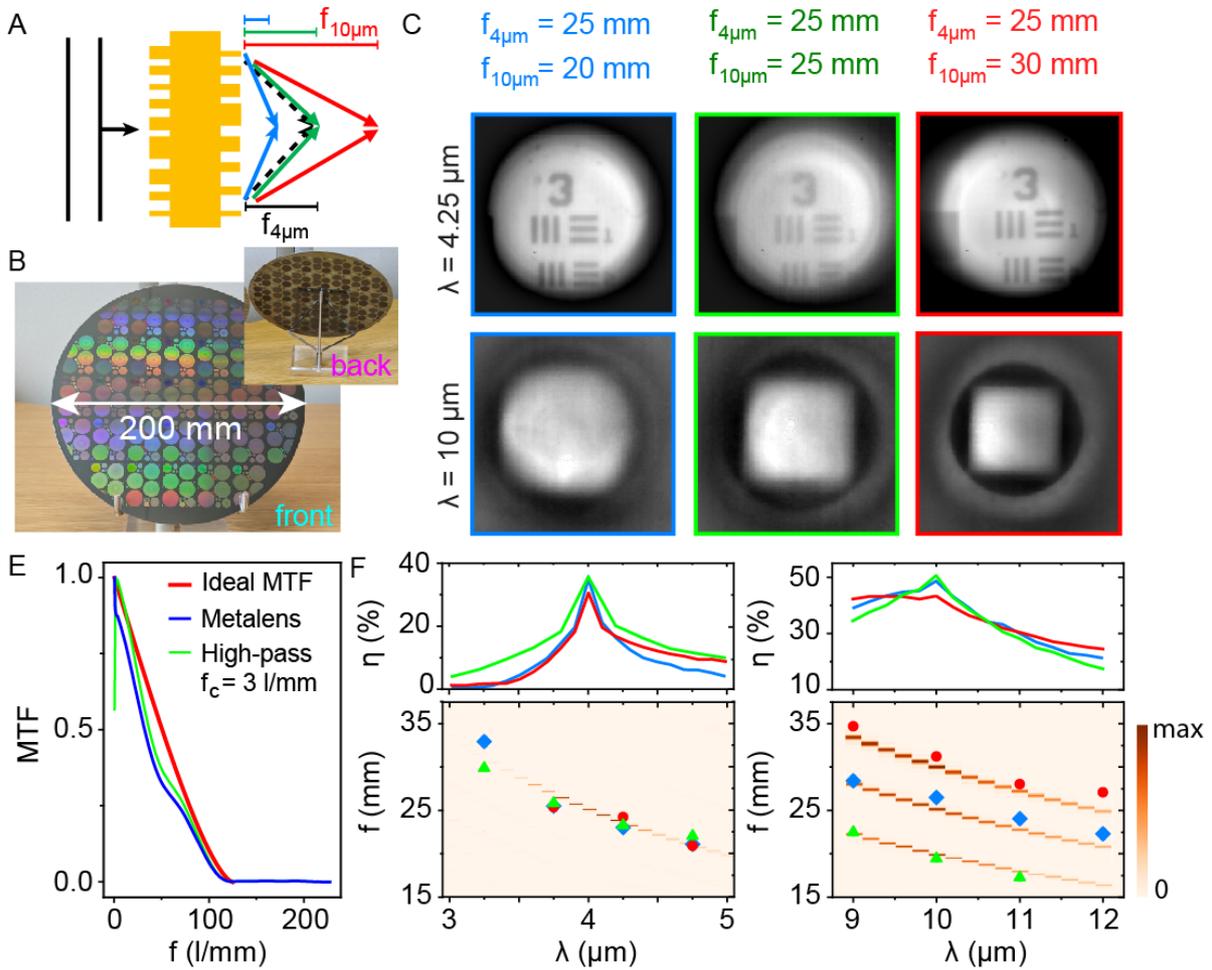

**Figure 3. Optical characterisations of the fabricated half-inch metalenses.** (a) Schematic illustration of concept of dual-band meta-optics with independent control over focal lengths in mid-wave and long-wave bands. (b) Photograph of front side of double-sided meta-optics wafer, with back side shown in inset. One wafer contains 110 individual lenses of 12.5 mm diameter. (c) Optical characterization of three metalenses with the same design focal length at λ = 4 µm and different focal lengths at long-wave λ = 10 µm using 1951 USAF test target (group 3) and MEMS light source as objects for imaging. (d) Calculated MTF for design metalens (blue) compared to ideal (aberration-free) MTF (red) at 4.0 µm. MTF correction using high-pass filtering at 3 lines/mm cutoff frequency to reduce haze background is shown by green curve. (e) Experimentally measured (symbols) focal distances compared to extracted intensities (colormaps) from simulations at mid-wave and long-wave bands. Top panels in (e) are simulated focus efficiencies η within a 50 µm diameter around the focus.

**2.4 Large-area double-sided meta-optics demonstrators**

Following the successful demonstration of double-sided meta-optics at 12.5 mm diameter, we proceed by increasing the meta-optic diameter to 40 mm. This increase in size requires a multi-exposure stitching approach, which was implemented using a single reticle per side as illustrated in **Figure 4**A, horizontal write fields are indicated by orange and vertical (rotated over 90°) are indicated by green colors. The reticle itself consists of two quadrants of the meta-optic in two opposing corners of the area (Section S2, Supporting Information). Accurate exposure is achieved by aligning with the predefined alignment markers at specific locations on the wafer, facilitating the alignment of exposure write fields after wafer rotation. In this case, the front-to-back alignment was also through the substrate notch but at the lithography steps for alignment mark definition on both sides. For 40 mm metalens, we have made two designs, one designed for MWIR/LWIR dual-band (V5 in Figure 1b) and the other one for LWIR single band (V6 in Figure 1b). Figure 4B shows the front and back sides of the resulting wafer (V5, dual-band) containing 9 individual large-area metalenses. The accuracy of multi-reticle stitching was found to be less than 100 nm as can be observed in Figure 4C showing a zoomed in SEM of the center of the meta-optic (V5). The central hollow square was segmented into four quadrants which are matched together well within the edge with of the hollow square, resulting in a well-defined composite structure with only very little underexposure at the edges resulting in very fine (<100 nm) protrusions around the stitched points.

Meta-optics components were laser-cut from the wafer and mounted into individual T-mount lens holders leaving 38 mm free aperture, as shown in Figure 4D. Then the imaging performance can be characterized by mounting the individual metalenses to the commercial MWIR and LWIR cameras as main focusing lenses. These individual meta-optics components can now be readily integrated into commercial MWIR and LWIR optical systems and used for real-world thermal imaging testing.

Results from benchtop tests are presented in Figure 4F-H and Figure 4K-M for a laboratory bench demonstration showing both the MWIR and LWIR performance on the same target object, which is a substrate with designed emissivity contrast in the form of the University of Southampton logo of approximately 5 cm in width positioned at 35 cm from the camera, with further details presented in Section S8, Supporting Information. The target is mounted on a hot plate and heated to 140°C. The availability of a 40 mm meta-optic allows for a true like-for-like comparison with state-of-the-art optical imaging lenses.

Figures 4F and K show the imaging results for the commercial lenses of our MWIR and LWIR cameras, respectively, which represent the 'ground truth' due to the superior build and achromatic performance of these multi-element lenses. For MWIR, a 2-inch diameter, 3.6 µm long-pass filter was placed in front of the lens to limit the response to the same range as the meta-optic. The LWIR reference image was cropped to an area of interest of F/1.5 (original image F/1) for better comparison with field of view of the meta-optics. The dual-band meta-optic shown in Figure 4G is able to produce a good quality image over its MWIR spectral band, with letters clearly identified, where some blurring and haze is expected based on the chromatic aberration of the lens. Results for the single-band lens are presented in Figure 4H. Clearly, the lack of MWIR functionality results in a complete loss of the MWIR image, which emphasizes the remarkable performance of the dual-band lens in comparison.

For the LWIR response, we use the freedom offered by the low F-number uncooled camera to improve the imaging quality by using a hybrid refractive and meta-optic lens pair. We make use of the fact that a $BaF_2$ lens shows the opposite sign of dispersion than the meta-optic, thus allowing a partial correction of chromatic aberrations whilst reducing the F-number as indicated in the configuration in Figure 4I and J. Here, the 25mm diameter $BaF_2$ lens is the smaller one of the pair and is located closest to the camera, at around 21 mm from the focal plane array (FPA) in Figure 4J. Resulting imaging results are presented in Figure 4L and M for the dual-band and LWIR meta-optic, respectively. We can clearly resolve the letters of the print as well as features of the shield logo down to 1 mm. In comparison, images generated using only the metalens (Figure S20, Section S9, Supporting Information) show a blurring by chromatic aberration. Additional short range imaging demonstrations are presented in Section S10, Supporting Information.

Further outdoor testing was done using the long-wave imaging system as shown in Figure 4N-P. The overall scene (Figure S23, Section S11, Supporting Information) shows features at distances of 10 m (person), 25 m (tree branches) and 150 m (roof line). Results for the commercial camera lens in Figure 4N can be directly compared with the results for the dual-band meta-optics (Figure 4O) and LWIR-only design (Figure 4P), both in hybrid $BaF_2$ refractive / meta-optic configuration. We note an improved performance for the single-band LWIR meta-optic over the dual-band lens as is expected given the trade-off in designing two separate operation wavelengths. In both cases the features at all three distances can be identified, verifying the operation of the meta-optic up to 150 meters range.

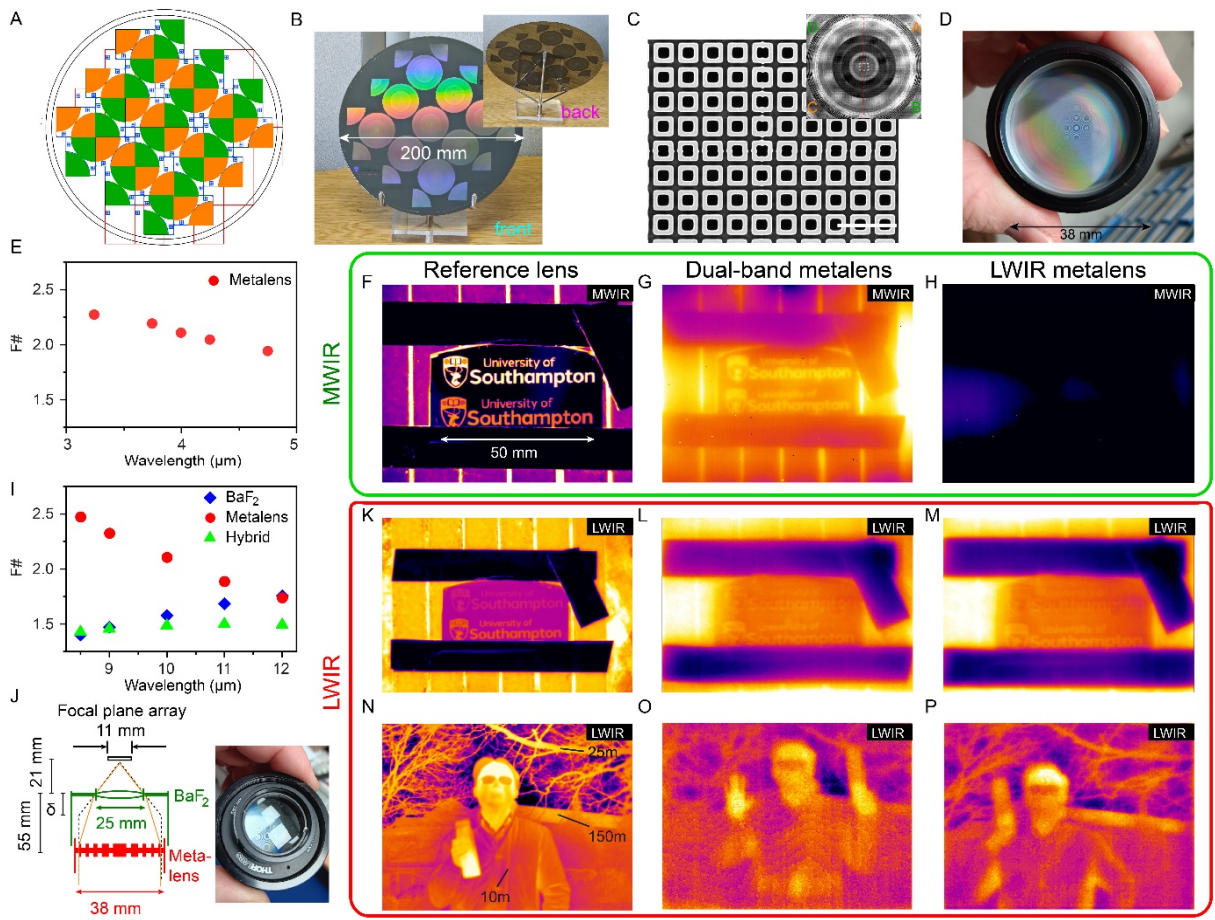

**Figure 4. 40 mm large diameter metalenses through multi-stitching exposure and their real-world thermal imaging performance.** (a) Layout of multi-exposure stitching with 90° wafer rotation using a single DUV reticle. (b) Photograph of front side of the produced double-sided meta-optics wafer containing 9 complete lenses of 40 mm diameter, inset shows back side of wafer. (c) SEM image of center of lens showing quality of stitched fields (scale bar, 5 µm), with inset showing a zoomed-out area indicating the different quadrants with color coded letters A-D according to exposures in (a). (d) Photograph of metalens laser-cut from the wafer and mounted in optical holder (T-mount) with 38 mm clear aperture. (e) F-number against wavelength for MWIR. (f-h) Direct comparison of metalens performance against state-of-the-art commercial lenses for mid-wave. (i) F-number against wavelength for LWIR. (j) Arrangement of hybrid metalens – refractive $BaF_2$ optical system. (j-l) Direct comparison of metalens performance against state-of-the-art commercial lenses for mid-wave. (f-h) and (j-l) obtained using target object heated to 140° C. (n-p) Outdoor imaging demonstration using hybrid $BaF_2$ and meta-optic compound at long-wave infrared, with details from 10 meters and up to 150 meters in range.

**Table 1.** Metalens Interferometry characterizations

|  | Dual-band metalens | | LWIR metalens |
|---|---|---|---|
| Wavelength | 3.39 µm | 9.24 µm | 9.24 µm |
| RMS wave-front error | 0.051λ | 0.042λ | 0.029λ |
| Peak-valley | 0.24λ | 0.28λ | 0.10λ |
| Strehl ratio | 0.90 | 0.93 | 0.97 |
| Astigmatism | 0.56λ at -18.1° | 0.06λ, at -87.1° | 0.04λ, at -63.3° |
| Coma | 1.02λ at 5.9° | 0.27λ at 153° | 0.24λ at -132° |
| Spherical aberration | 0.06λ | 0.30λ | 0.26λ |

The large-area double-sided meta-optics were further characterized using a range of industrial methods commonly used for refractive optical systems with further details in Section S13, Supporting Information. Results are summarised in Table 1 and include root mean square (RMS) wavefront error, peak-valley ratio (PV), Strehl ratio, astigmatism, coma and spherical aberration at wavelengths of 3.39 µm and 9.24 µm for MWIR and LWIR, respectively, which were chosen based on available laser sources. For the dual-band metalens, its MWIR performance is characterized by a low Peak-valley of 0.24 λ and high Strehl ratio of 0.90, which exceed the typical threshold values of <0.25 λ for Peak-valley and >0.80 Strehl ratio and therefore indicate that this dual-band metalens is diffraction-limited, that is, its optical performance is governed by fundamental physics rather than imperfections. The RMS wave front error of 0.051 λ also shows the metalens has minor residual imperfections, which could be from the astigmatism, coma and spherical aberration. Its LWIR performance also results in a low RMS of 0.042 λ, low Peak-valley of 0.28 λ and high Strehl ratio of 0.93, which are also diffraction-limited. Compared with the dual-band metalens at 9.24 µm, the LWIR metalens gives a lower RMS, lower PV and higher Strehl ratio and also lower astigmatism, coma and spherical aberration. The observation of superior values for the LWIR metalens well matches the seen superior performance of the LWIR metalens over the dual-band metalens in real world imaging (Figure 4P and O).

3. Conclusion

In conclusion, we report a novel method for manufacturing wafer-scale double-sided meta-optics with full compatibility with an automatic DUV scanning lithography system and Track. Based on this manufacturing method, we demonstrate dual-band MWIR/LWIR metalenses with independent focal length controls. In addition, we have further demonstrated that 40 mm large-diameter double-sided metalenses with dual-band and single-band operations can be fabricated through a multi-exposure stitching method to overcome the DUV exposure field limit. The fabricated metalens performance was demonstrated through their integration into imaging systems for real-world thermal imaging. Interferometry measurements show that these metalens are diffraction-limited with minimal aberrations and wavefront imperfections.

The proposed double-side meta-optic manufacturing method opens up a new route to infrared meta-optics designs, and scale-up production of double-side meta-optics manufacturing. The double-sided meta-optics offers an expanded design space for multi-functionality, such as the dual-band presented in this work, or could accommodate moth-eye anti-reflection coatings, doublets or aberration meta-correctors. The demonstrated double-sided lithography technique is compatible with other scale-up manufacturing techniques for meta-optics, such as nanoimprint lithography and direct laser writing, which can lead to novel designs and structures beyond existing capabilities.

4. Experimental Section/Methods

*Metalens design*: Double-sided metalenses were designed using RSoft Photonic Design Tools (Synopsys) on a dual-processor Xeon Gold 6148 2.40GHz workstation with 512 GB RAM. The unit cell was designed using the RSoft CAD environment, subsequently the bidirectional scattering distribution (BSDF) function was simulated using rigorous coupled wave analysis (RCWA) implemented in the DiffractMOD simulation tool, taking advantage of parallel processing over 32 nodes in the multi-variable optimization and scanning tool (MOST). A range of incident angles from 0 - 60° with steps of 15°, wavelengths from 3.0 – 5.0 µm in steps of 0.1 µm for MWIR and 9.0 – 12.0 µm in steps of 0.2 µm for LWIR. The unit cell was modelled using four rectangular silicon pillars equal in size, each pillar with a long axis length ranging from 0.2 – 1.8 µm and short axis width ranging from 0.2 – 0.9 µm. The pillars were arranged to form a square or cross shape, depending on the offset parameter and the length to width ratio as is illustrated in Figure 2. The unit cell period was varied in a series of simulations from 1.6 – 2.6 µm and an optimal period of 2.2 µm was found to give best overall performance. The vertical height of the pillar structures was set to 3.5 µm for the dual-band design as defined by the achievable etch depth in our process. The period and height for the LWIR only design was chosen as 2.6 µm and 4.0 µm, respectively, following further optimization for the 10 µm wavelength design in absence of a MWIR design target and slight improvements of the etch depth made in the experimental process development. Metalenses were designed from the generated BSDF database using the MetaOptics Designer simulation tool, using an on-axis Airy spot focus as the target objective. An intensity-difference metric was chosen with a weighting which was strongly biased towards achieving a high absolute transmission efficiency with less weight given to matching the exact shape of the Airy profile. The simulation used absorbing boundary conditions and taking into account local propagation angles inside the double-sided metalens structure. These parameters were found to yield convergence toward near-diffraction limited spots with corresponding good-quality MTF and an absolute focusing efficiency limited by the available degrees of freedom in the database. A GDS-II output of both sides of the metalens was generated and was further processed using KLayout to include alignment markers, SEM bars, front-to-back misalignment scale, and several other test structures. Deep-UV reticles were manufactured through MacDermid Alpha Electronic Solutions.

*Deep-UV lithography fabrication of meta-optics*: Here, the fabrication process flow is for the 40 mm metalens with exposure stitching. The fabrication of metalens started by growing 800 nm $SiO_2$ on double polished 200 mm Si substrates through wet oxidation at 1000 °C. Through a Nikon NSR-S204B DUV scanning system and TEL Track (an automated resist coating and developing system), the alignment mark was defined on the front side of Si substrates and subsequently transferred into the $SiO_2$ layer through a plasma etching. Subsequently, the alignment marks were also defined on the back side of Si substrates through the same DUV and plasma etching, with the DUV front-to-back alignment done through notch recognition. Aligning to the defined alignment marks, DUVsteps, with and without substrate rotation of 90°. The pattern was transferred into the hard mask $SiO_2$ through an ICP plasma etching using fluorine chemistry. After the photoresist removal, the back side $SiO_2$ hard mask was also processed through the same DUV lithography process and plasma etching process. After the initial preparation of the $SiO_2$ hard masks on both sides of the wafer, the Si pillar arrays were transferred from the $SiO_2$ arrays using an ICP plasma etch with $SF_6$ and $C_4F_8$, first on one side and subsequently on the other side. The etch depth was controlled through etching time with SEM images in Figure S3, Section S3, Supporting Information. After the etching, an $O_2$ plasma treatment was given to remove the polymer formed during the Si etching. Finally, the $SiO_2$ hard masks on both sides of the wafer were stripped in 7:1 buffered hydrofluoric acid (HF).

*Infrared measurements*: Broadband response was measured using a blackbody light source (Axetris EMIRS200) with a 1.8×2.1 $mm^2$ active area. Measurements of short focal distance metalenses were done in reverse configuration, where the metalens was used as the imaging objective of a target object. The target was either the EMIRS light source itself or a 1951 USAF target. An achromatic imaging lens matching each specific camera was used to project the collected image onto the camera. For the MWIR range, we used a CEDIP/FLIR Titanium SC7300 camera with 320×256 pixels and a pixel pitch of 30 µm. The matching camera lens for the MWIR was a 50 mm focal length Janos Technologies Nyctea F/2.3 lens (Model 40679, wavelength 1.5 – 5 µm). For the benchtop demonstrator we used a 3.6 µm, 50 mm diameter, infrared long-pass filter (Edmund Optics) to limit the spectral range. For the LWIR range, we used a FLIR A655sc uncooled camera with 640×480 pixels and pixel pitch of 17 µm. The matching camera lens was a 24.6 mm focal length F/1 FLIR lens (Model T197922, 25° FOV). Diagrams of optical setups used for the focal length characterization are shown in Figure S16, Section S7, Supporting Information.

*Quantitative testing of metalens performance*: The MWIR performance in Table 1 was measured by a 3.39 µm wavelength phase-shifting Michelson Interferometer with a default aperture size of 50 mm. The LWIR performance was measured by a longwave Twyman-Green interferometer with a default aperture size of 250 mm, equipped with a $CO_2$ gas laser at 9.24 µm. A discussion of these tests is presented in Section S13, Supporting Information.

Acknowledgements

The authors acknowledge financial support by the EPSRC National Hub in High Value Photonic Manufacturing (EP/N00762X/1). JS acknowledges support from an EPSRC collaborative PhD studentship with Teledyne Qioptiq.

Conflict of Interest

The authors, K.S. and O.L.M., have filed a patent on the double-sided metasurface manufacture technique with United Kingdom Patent Application No. 2503168.3, "Optical Metasurfaces and Method of Fabrication of Optical Metasurfaces". JNM is employed by Teledyne Qioptiq Ltd.

**Uncategorized References**

**Supporting Information: Production-ready double-side fabrication of dual-band infrared meta-optics using deep-UV lithography**

*Kai Sun\*, Xingzhao Yan, Jordan Scott, Jun-Yu Ou, James N. Monks, Otto L. Muskens\**

Kai Sun, Jordan Scott, Jun Y. Ou, Otto L. Muskens
Physics and Astronomy, Faculty of Engineering and Physical Sciences, University of Southampton, Southampton, SO17 1BJ, United Kingdom
E-mail: K.Sun@soton.ac.uk, O.Muskens@soton.ac.uk

Xingzhao Yan
Optoelectronics Research Center, Faculty of Engineering and Physical Sciences, University of Southampton, Southampton, SO17 1BJ, United Kingdom

James N. Monks
Teledyne Qioptiq ltd, Glascoed Rd, Saint Asaph, LL17 0LL, United Kingdom

# S1. Overview of selected publications on metalens fabrication with focus on large-area, dual-band and infrared range

**Table S1 Summarized selected metalens works since 2020.**

| Year | Band | Diameter (mm) | Double-side | Multi-band | Lithography | Material | Substrate | Substrate size | Ref |
|---|---|---|---|---|---|---|---|---|---|
| 2025 | LWIR | 24, 32 | | | Laser | Si | Si | | [1] |
| 2025 | Vis. | 2.8 | | | EBL | a-Si | Glass | | [2] |
| 2025 | LWIR | 20 | | 2-WL | Laser | Si | Si | | [3] |
| 2025 | LWIR | 31.6, 63.6 | | | | c-Si | Si | 4 | [4] |
| 2024 | Vis. | 0.5 | | | EBL | a-Si | Glass | | [5] |
| 2024 | Vis. | 0.6 | | | Laser | Graphene-oxide | Glass | | [6] |
| 2024 | MWIR | 11.95 | Yes | | EBL | Si | Si | | [7] |
| 2024 | LWIR | 12.2 | | | DUV | Ge | Ge | 2 | [8] |
| 2024 | LWIR | 10 | | | Laser | Si | Si | | [9] |
| 2024 | NIR | 2.5 | | | EBL | a-Si | Glass | | [10] |
| 2024 | LWIR | 100 | | | Laser | Si | Si | | [11] |
| 2024 | LWIR | 40 | | | Laser | Si | Si | | [12] |
| 2024 | LWIR | 6 | | | Laser | Si | Si | 4 | [13] |
| 2024 | Vis | 0.6 | | | 2-photon | TiO2 | Glass | | [14] |
| 2024 | NIR | 1.58 | | | EBL | a-Si | Glass | | [15] |
| 2024 | NIR | 10 | | | NIL | a-Si:H | Quartz | 4 | [16] |
| 2024 | Vis. | 100 | | | DUV | $SiO_2$ | Glass | 6 | [17] |
| 2024 | Vis. | 5 | | | DUV | a-Si | Glass | 12 | [18] |
| 2024 | NIR | 4 | | | EBL | a-Si | Glass | | [19] |
| 2024 | Vis. | 0.1 | | | 2-photon | Polymer | Glass | | [20] |
| 2024 | LWIR | 6 | | | DUV | Si | Si | | [21] |
| 2024 | SWIR | 1.28 | | | EBL | c-Si | Sapphire | | [22] |
| 2023 | Vis. | 0.5 | | 3-WL | EBL | a-Si:H | Glass | | [23] |
| 2023 | Vis. | 0.2 | | | EBL | $TiO_2$ | Glass | | [24] |
| 2023 | Vis. | 0.16 | | | EBL | Si | Sapphire | | [25] |
| 2023 | Vis. | 10 | | | DUV | Polymer | Glass | | [26] |
| 2023 | Vis. | 6 | | | NIL | SiN | Glass | | [27] |
| 2023 | SWIR | 0.4 | | | EBL | Si | Si | | [28] |
| 2023 | SWIR | 0.03* | Yes | | EBL | Si | Si | | [29] |
| 2023 | LWIR | 20* | | 2-WL | laser | Si | Si | 2 | [30] |
| 2023 | SWIR | 80 | | | DUV | a-Si | Glass | 4 | [31] |
| 2022 | Vis. | 10 | Yes | | EBL | Si, SiN | Sapphire | | [32] |
| 2022 | SWIR | 0.5 | Yes | | EBL | a-Si:H | Glass | | [33] |
| 2022 | LWIR | 80 | | | Contact | Si | Si | 4 | [34] |
| 2022 | SWIR | 0.04 | | | EBL | Si | SOI | | [35] |
| 2021 | Vis./SWIR | 0.2 | | Yes | 2-photon | Nanoholes | Glass | | [36] |
| 2021 | LWIR | 20 | | | laser | Si | Si | 4 | [37] |
| 2021 | MWIR | 0.7 | | | DUV | Ge | Sapphire | | [38] |
| 2021 | Vis. | 0.5 | | | EBL | a-Si | Glass | | [39] |
| 2021 | SWIR | 0.6 | Yes | | EBL | a-Si | Glass | | [40] |
| 2020 | NIR | | Yes | | EBL | a-Si | Glass | | [41] |
| 2020 | Vis./SWIR | 0.01-0.02 | | Yes | EBL | $TiO_2$ | Glass | | [42] |
| **2025** | **MWIR/LWIR** | **40** | **Yes** | **Yes** | **DUV** | **Si** | **Si** | **8** | **This work** |
| **2025** | **LWIR** | **40** | **Yes** | | **DUV** | **Si** | **Si** | **8** | **This work** |

*: square shape, WL: wavelength

Glass: Glass stands for fused silica, quartz and glass substrates, a-Si: amorphous silicon, Si: crystalline silicon.

Vis.: visible, NIR: near-infrared, SWIR: short-wave infrared, MWIR: mid-wave infrared, LWIR: long-wave infrared, DUV: deep ultraviolet lithography, EBL: electron beam lithography, 2-photo: two-photon lithography, Contact: contact lithography, NIL: Nanoimprint lithography.

## S2. Multi-exposure stitching

Two reticles were designed with one for front and the other one for back side, as shown in Figure S1. For the front side labelled A, it includes two quarters of a designed 40 mm diameter metalens layout. Ideally, one quarter can be used to achieve full lens through rotation. However, our DUV is configured with a capability of one 90-degree rotation. Here, it is noted that the DUV reticle can only be loaded with one orientation as there are extra patterns on the reticle edge for machine identification. Thus, two quarters are needed to achieve a full-round lens layout. In addition, there are also misalignment measurement patterns and two alignment mark patterns on top-right and bottom-left, which are used to define alignment marks for metalens exposures and its rotation exposures. For the back side labelled B, the metalens patterns are in similar layouts but mirrored.

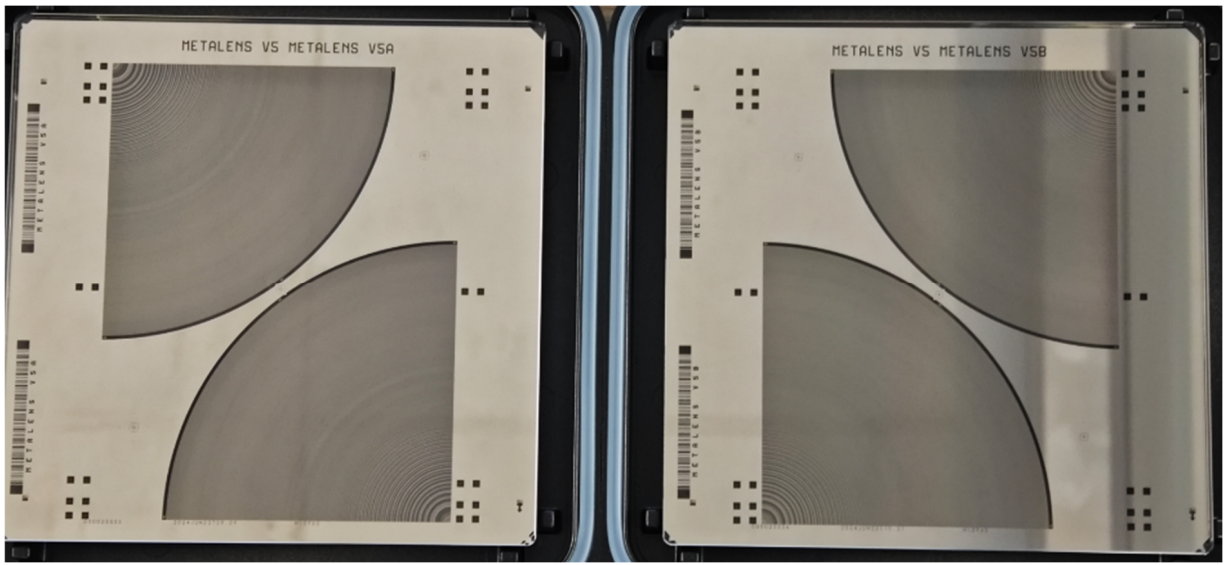

**Figure S1 Reticle layouts for multi-exposure metalens designs (40 mm in diameter) (a) front side and (b) back side.**

The multi-exposure strategy is illustrated in Figure S2 involving six lithography steps, using the two reticles and substrate rotation functions. Firstly, the alignment mark is defined on the substrate Front side at edge regions as illustrated (Figure S2(a)) where no metalens will be located on. Secondly, the alignment mark is defined on substrate back side in a mirrored style (Figure S2(b)), with alignment through wafer notch. Thirdly, metalens layouts (two quarters) are defined at defined locations on the front side (Figure S2(c)), with alignment control to predefined alignment marks. Fourthly, the other two quarters of metalens layouts are defined using the same reticle but at wafer rotation of 90º, with alignment control also to predefined alignment marks (Figure S2(d)). Fifthly, the back side layouts are processed in the same manner, with two quarters defined on back side using back side alignment mark (Figure S2(e)). Sixthly, the rest two quarters are defined through substrate rotation by 90º (Figure S2(f)). Here, the design has to be mirrored to reflect the opposite relevant rotation orientation on different sides.

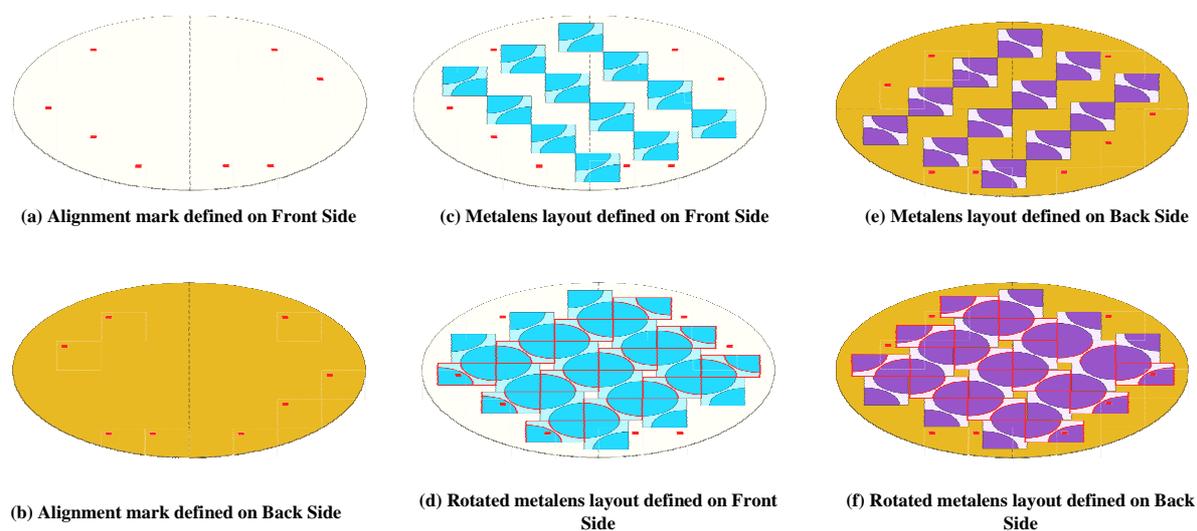

(a) Alignment mark defined on Front Side

(c) Metalens layout defined on Front Side

(e) Metalens layout defined on Back Side

(b) Alignment mark defined on Back Side

(d) Rotated metalens layout defined on Front Side

(f) Rotated metalens layout defined on Back Side

**Figure S2 Schematic of multi-exposure stitching process flow. (a) Alignment marks defined on front side of the substrate, (b) Alignment marks defined on back side of the substrate, (c) metalens layout partially exposed on front side and (d) metalens layout partially on the front side with 90o substrate rotation, (e) metalens layout partially exposed on back side, and (f) metalens layout partially exposed on back side with 90o substrate rotation.**

### S3. Scanning electron microscopy images

Figure S3 presents the cross-sectional SEM images of one etch test wafer with the $SiO_2$ hard mask on Si pillars. For the front side in Figure S3(a), the etch depth varies from 3.24 to 3.64 µm with a mean etch of 3.44 µm. Si pillars (square rings) are seen to be well-defined and vertical. The $SiO_2$ hard mask layer is about 379 nm left. For the back side in Figure S3(b), the etch depth varies from 3.27 to 3.74 µm with a mean etch of 3.50 µm. Si pillars (square rings) are seen to be well-defined and vertical. The $SiO_2$ hard mask layer is about 390 nm left. Therefore, etch depth on both sides can be well defined through timed etching.

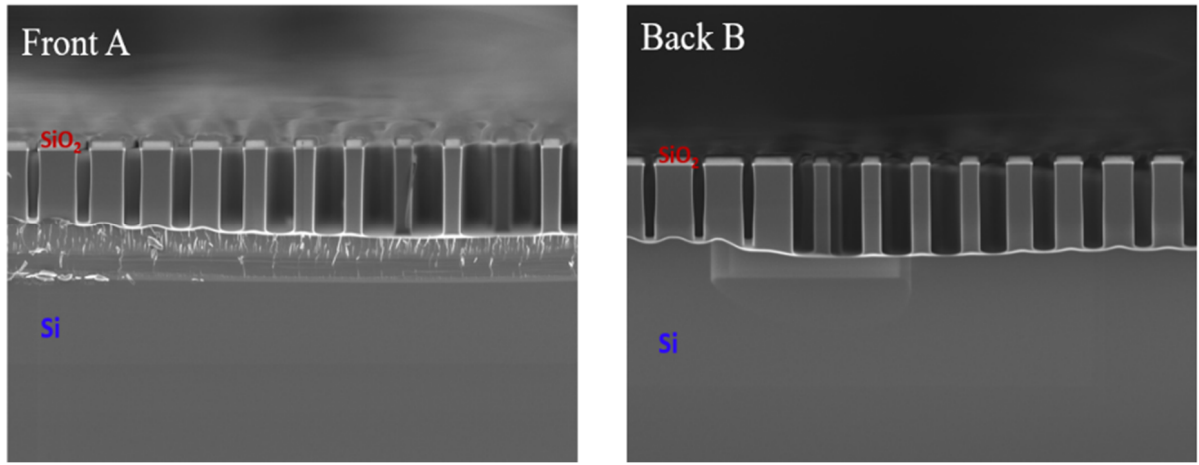

**Figure S3 Cross-sectional SEM images of front side (A) and back side (B) of etched Si metalens structures on an etch test wafer for design V5, cleaved through the hollow squares. Etch depth variations with gap size can be seen due to microloading effect.**

SEM inspections were performed on a metalens as presented in Figure S4, which is from the test wafer given the same etch conditions, with $SiO_2$ removed but no $O_2$ plasma treatment given. From these images, the square rings and square pillar are well defined. The central stitching can be identified, indicating some misalignments at the bottom two quadrants in x-axis in a few ten nm but superb for the top two quadrants. There is some ring deformation at the stitching joints and this is attributed to the slight lower exposure at the very edge of the exposure field. In future work, these 'joints' between individual rings can be corrected by slightly modifying the features across the stitching boundaries or shift the feature out of the stitching boundaries. However, the impact is likely to be quite limited.

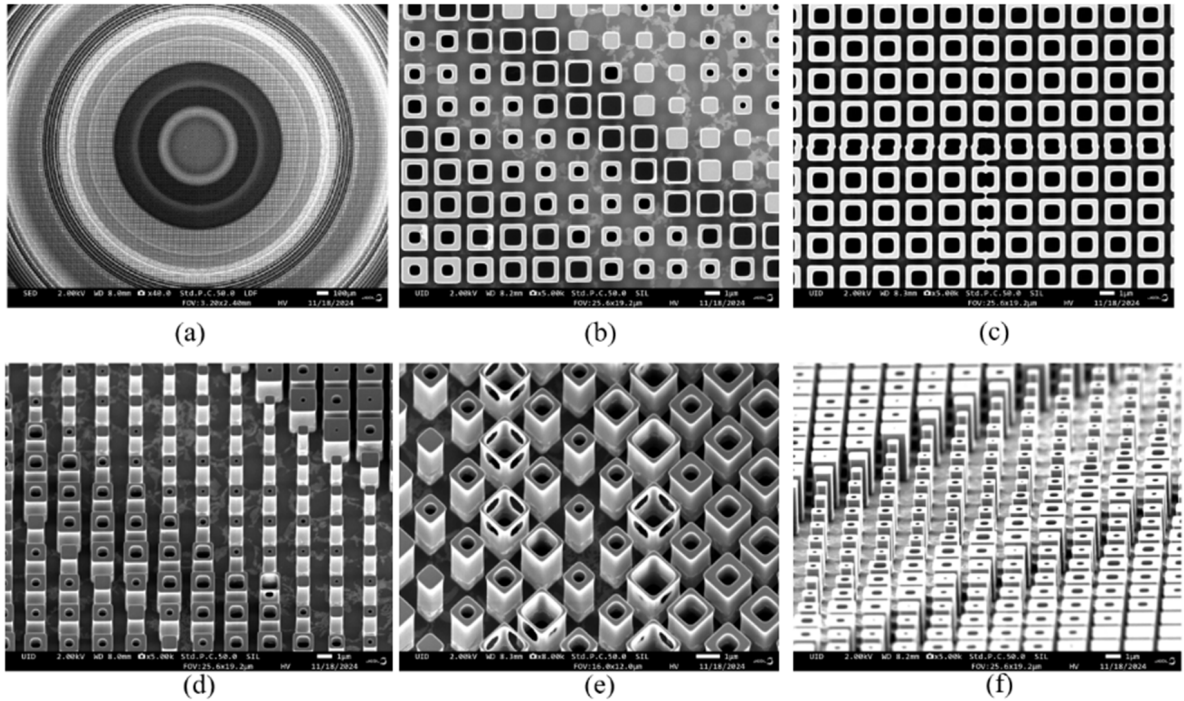

**Figure S4** SEM images of the Front side of a 40 mm metalens wafer V5, (a) low-magnification at the metalens center, (b) top-right near-the edge, (c) stitching at the center, (d) 25° tilt, (e) 25° tilt and 25° rotation, and (f) 65° tilt. SEM bars are 100 μm for (a), 1 μm for (b) and 1 μm for (c)-(f).

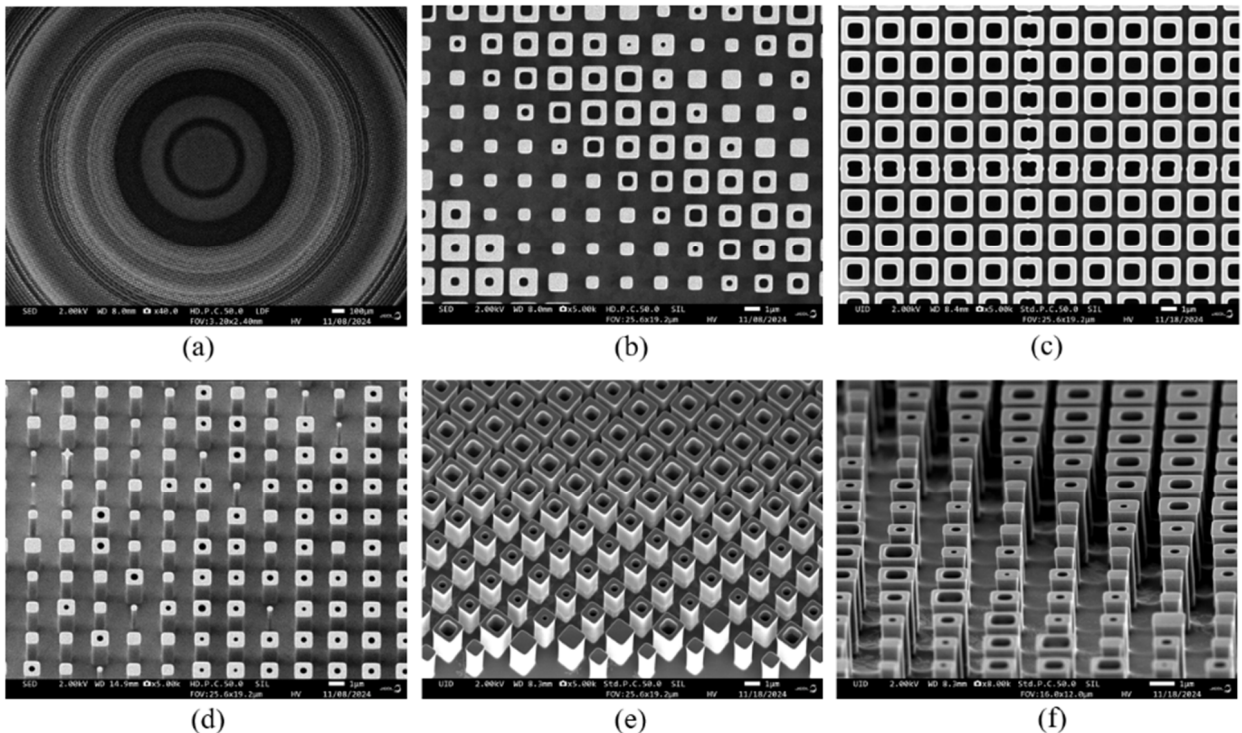

**Figure S5** SEM images of the Back side of a 40 mm metalens (Batch 3), (a) low-magnification at the metalens center, (b) top-right near-the edge, (c) stitching at the center, (d) 25° tilt, (e) 25° tilt and 25° rotation, and (f) 65° tilt. SEM bars are 100 μm for (a), 1 μm for (b) and 1 μm for (c)-(f).

The images give further insight in the formed structures with smooth sidewalls. It is of interest to see that in some cases holes are formed at the sidewalls of the thinnest square rings, which can be attributed to plasma ion deflection during the etching. In general, the Si etching control is believed to be excellent. The results for the back side shown in Figure S5 are similar to those on front side. In summary, the nanostructures are well-defined including square pillars, cross pillars to square ring pillars, with excellent stitching control.

**S4. Pattern registration verification on both sides of wafer**

One of the key technical challenges is to align the two metalens sides without dedicated alignment marks as the DUV system is not configured for back side alignment. Instead, the alignment was achieved through wafer notch identification using its own automatic system. We have set alignment measurement patterns on both reticle layers at the center of exposure field as in Figure S6(a). It has a vernier scale for 4 µm misalignments (4 units) on both X and Y directions. For misalignment above 4 µm, it can be estimated directly through the center pattern shifts. The misalignment measurement was done through our MWIR camera as the silicon substrate is transparent in this range. As the wafer is manually mounted into the camera system some global rotation of the marker fields is observed in most of the images which does not relate to the front-back misalignment. Figure S6(b) and (c) shows typical IR images of misalignment patterns taken at focus of Front A and Back B, respectively. From Figure S6(b), we can estimate a misalignment of about 21 µm and 4 µm in X and Y directions, respectively

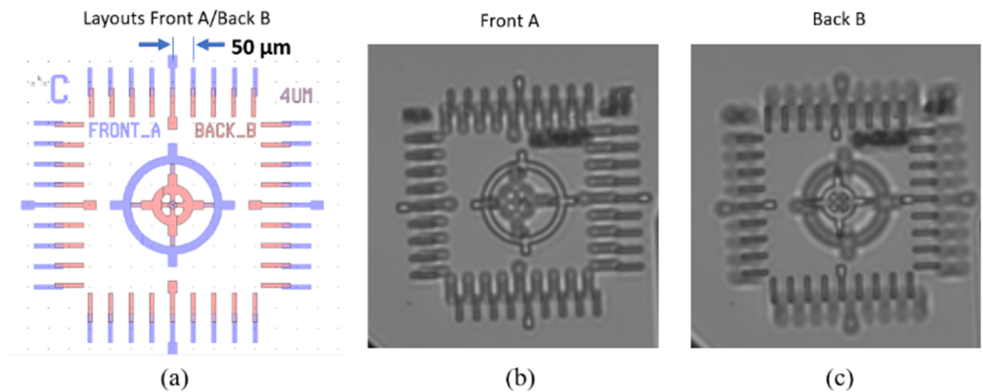

**Figure S6 Front-back side misalignment inspection, (a) designed layouts, (b) micrograph focused on Front A side and (c) micrograph focused on B side.**

Figure S7 shows misalignment measurements at different locations over the wafer V4. We can see the Back B side shifts towards right in x-direction at the top region and shifts towards left direction in X at the bottom region, and the shifts only little in y-direction. The same trend is also seen in two more wafers, indicating some consistent patterns. Therefore, we conclude that the misalignment might result from a rotation of approximately 0.045 °.

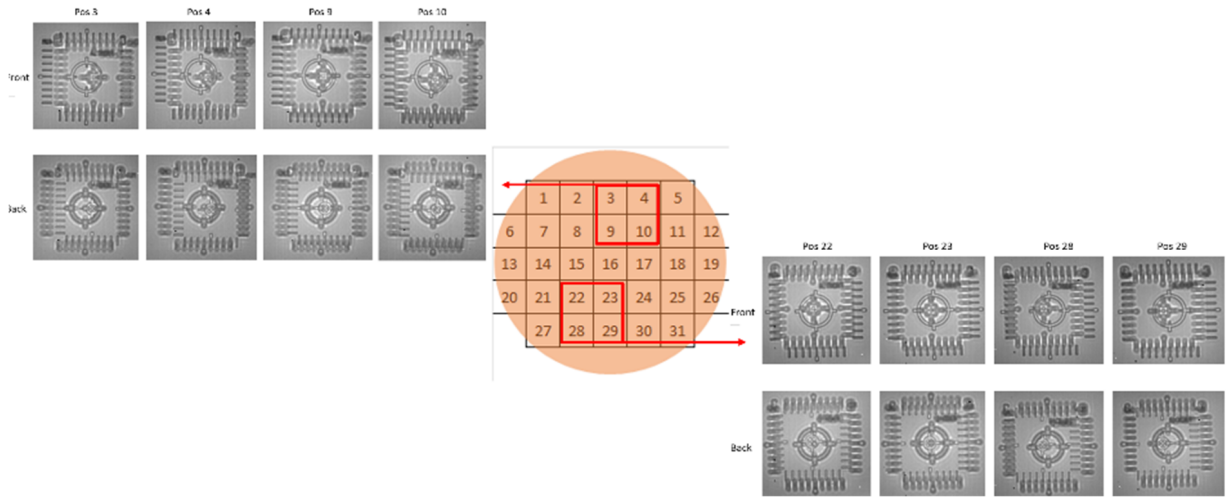

**Figure S7 Front-back side misalignment inspection for eight different locations on the wafer V4. Front / back indicates optical micrograph focusing on corresponding side of the wafer.**

We hypothesize that this rotational misalignment could be systematic either in the tool calibration or in the notch shape when mirrored as the wafer flips over. To investigate whether better alignment can be achieved through a rotation offset, we applied this correction (in both clockwise and counterclockwise directions) to two wafers in the V5 batch. We developed two wafers with opposing wafer rotations of 0.045 degrees in the second exposure. Figure 58 shows results for wafers without correction (#15), and with rotational corrections for +0.045° (#16) and -0.045° (#17). We tested the rotation correction in both directions, in order to avoid any ambiguity in the direction of the rotation correction. Markers on four positions of the wafer were analyzed corresponding to top left, bottom left, bottom right and top right.

The misalignment can be most easily read out by looking at the relative positions of the two front and back cross-hairs in the center of the alignment markers. For a good alignment, the cross-hairs line up in the center as in case of Wafer #17 top left location. We see that for Wafer #16 the misalignment is larger than for the uncorrected wafer, indicating that in this case the correction was applied in the wrong direction. For Wafer #17 it appears that the correction has partially worked and the error in misalignment has been reduced compared to the uncorrected wafer.

From this study we conclude that it is possible to correct for systematic misalignment in the system once the error has been identified. We note that each notch is laser marked by the wafer supplier to identify a front side and we can therefore keep track of the orientation of all wafers going through the system. So far, we have used one type of wafers from one supplier, and we cannot exclude batch variations. Therefore, it will be important to keep track of alignment accuracy going forward to establish a full understanding of this parameter between fabrication runs, when changing wafer batch, or even after DUV tool servicing.

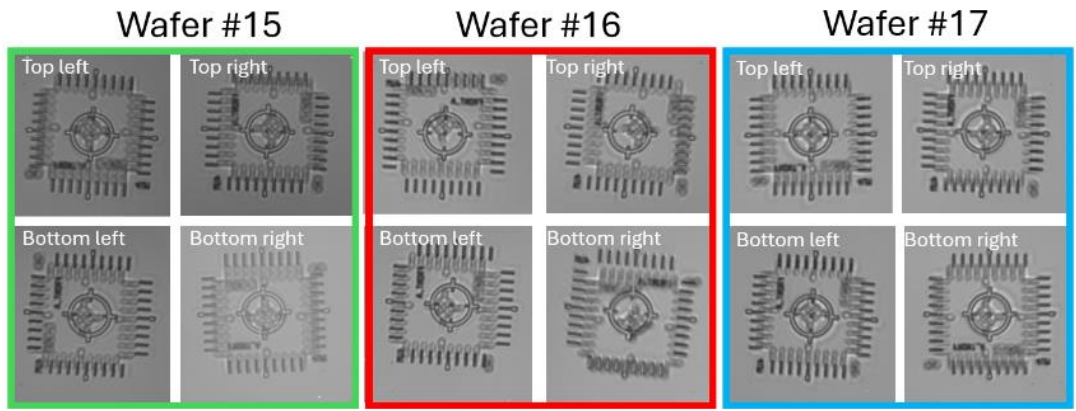

**Figure S8 Wafer front-back side misalignment inspection for three different wafers V5 without correction (#15) and with rotational corrections for +0.045°(#16) and -0.045°(#17).**

## S5. Large-area double-sided metalens V5 and V6 – design and characterization

To compare the results for front and back side for the probability of selection for each element in the design database, results for both sides are shown in Figure S9. Both sides have similar distributions, indicating that the simulation model has converged to a similar choice of elements on each side.

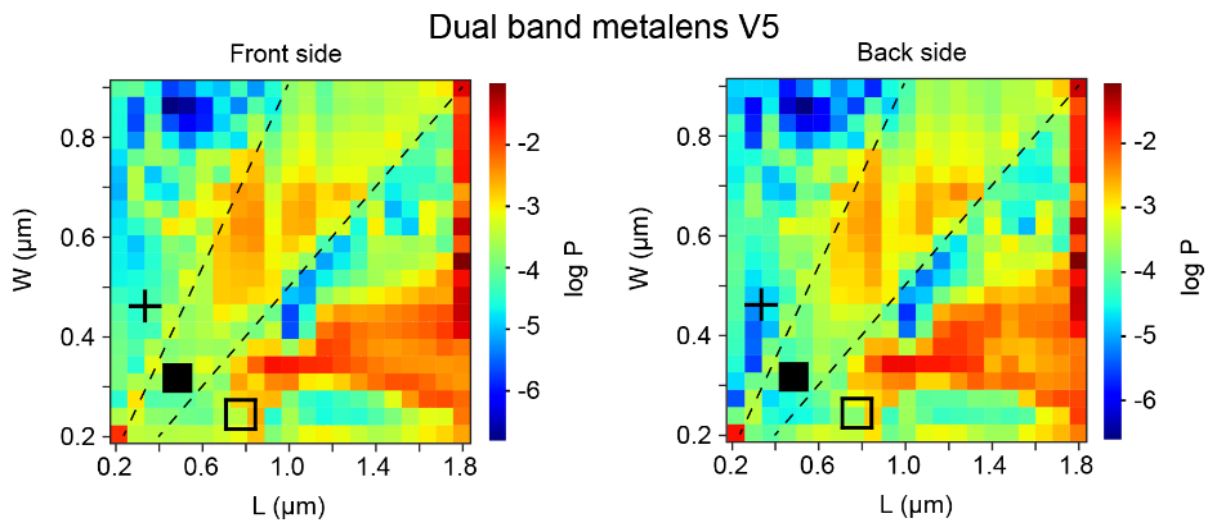

Figure S9 Probability of selection of invidual elements in the design space of the large-area 40 mm diameter, dual-band metalens V5, for front side and back side.

For the single-band LWIR metalens design V6, the absence of constraints for MWIR performance allows for a revisiting of the design space as presented in Figure S10. We tested different periods and found that increasing the period to 2.6 µm resulted in slight improvement of performance compared to the 2.2 µm period used in the dual-band design. Otherwise a similar databased model was chosen consisting of hollow squares, filled squares and crosses, where the length and width parameters were varied from 0.2 – 2.3 µm and 0.2 – 1.0 µm respectively. Following onto slight improvements in the experimental process optimization in the previous fabrication runs, we increase the achievable etch depth to 4.0 µm, which allows achieving a larger range of optical phase values. Figure S10A, B show the transmission and phase of the database for a single metasurface.

Using this database, we design the LWIR double-sided metalens V6 in MetaOptics Designer, using the same parameters as the dual-band but with only the single-wavelength optimization target. Resulting probability maps for the front and back side are shown in Figure S10C, D which shows that the model convergence on a number of well-defined bands within the 2D parameter space.

The simulated designs for the large-area double-sided metalens demonstrators are summarized in Table S2, showing the calculated AFE values in both bands. An efficiency increase up to 69.6% at LWIR is obtained by using the single-band design compared to the dual-band metalens with design performance at 49.9% AFE.

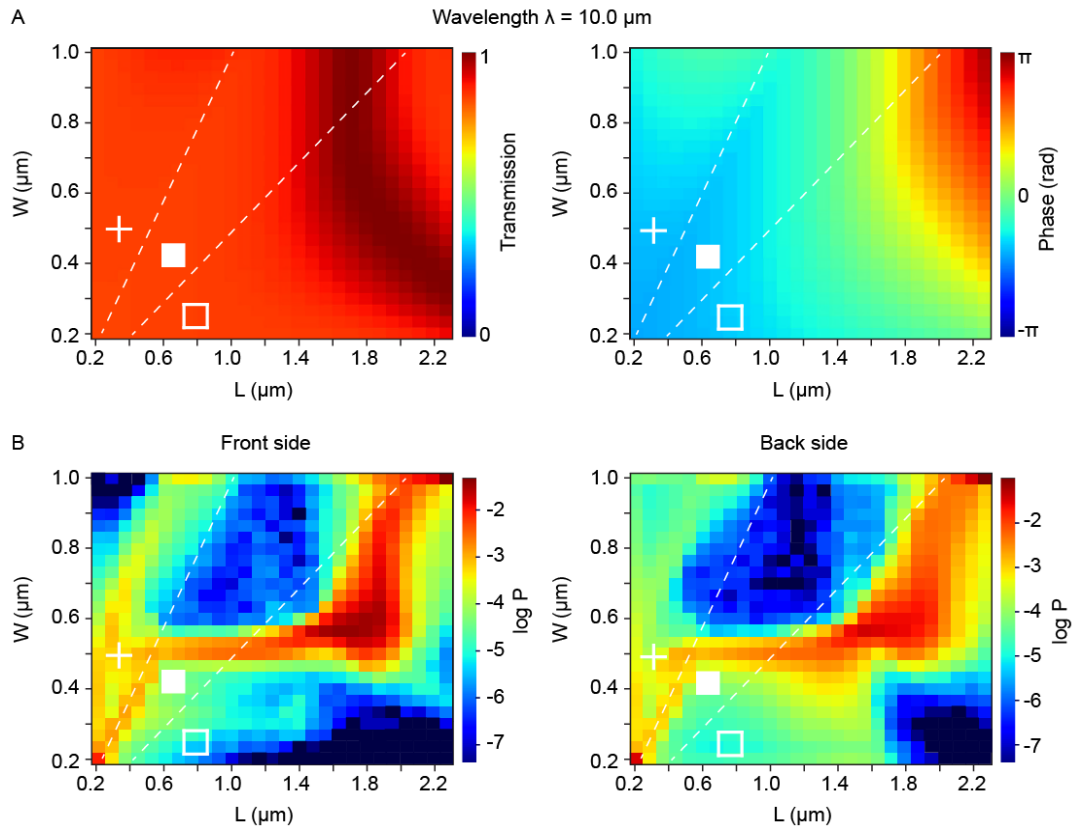

**Figure S10 Design space of double-side LWIR metalens. A** Transmission and Phase of the database at design wavelength of 10 μm. **B** Probability of selection of individual elements in the design space of the large-area 40 mm diameter, dual-band metalens V6, for front side and back side.

**Table S2 Design parameters for the large-area double-sided metalens demonstrators.**

| Wavelengths (μm) | Focal length @4μm (mm) | Focal length @10μm (mm) | Height (μm) | Diameter (mm) | $\eta_{4\mu m}$ | $\eta_{10\mu m}$ |
|---|---|---|---|---|---|---|
| 4/10 (V5) | 80 | 80 | 3.5 | 40 | 28.4% | 49.6% |
| 10 (V6) | - | 80 | 4.0 | 40 | - | 69.6% |

## S6. High-pass image de-hazing filter

The process of image enhancement by applying a high-pass enhancement filter is explained in Figure S11. The filter removes characteristic haze of the metalens image below 3 lines / mm and produces a better contrast with markedly more visible background features. The process is implemented using Python as illustrated in Figure S11B and C. First, a high-pass filtered image is generated by subtracting a 30-pixel Gaussian low-pass filter from the original image, resulting in the high-pass image as shown in Figure S11B containing both negative and positive contributions. The high-pass enhanced image is shown in Figure S11C and is obtained by adding the original image to the high-pass image. The processing ensures that the original image is retained in the case of a flat background (no high-pass components) whereas features with spatial frequencies higher than 3 lines / mm are sharpened.

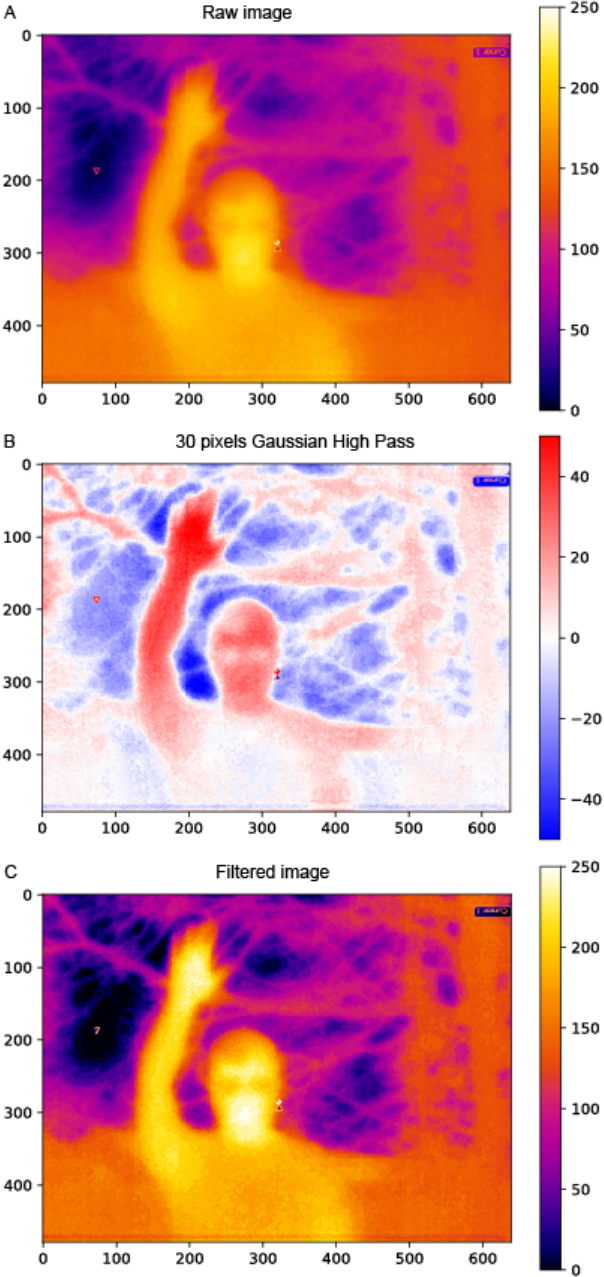

**Figure S11 Details of image processing using a 30 pixels Gaussian filtering to reduce blurring and haze below 3 lines/mm spatial frequency.**

## S7. Dual-band double-sided metalens V4 – design and characterization

The details of the three dual-band double-sided metalens designs, for demonstrating independent control over focal length, are shown in Table S2. All lenses are 12.5 mm in diameter and with a design etch depth of 3.5 µm.

**Table S3 Three dual-band metalens designs with different focal lengths at 4 µm and 10 µm wavelength.**

| Wavelengths (µm) | Focal length @4µm (mm) | Focal length @10µm (mm) | Height (µm) | Diameter (mm) | $\eta_{4\mu m}$ | $\eta_{10\mu m}$ |
|---|---|---|---|---|---|---|
| 4/10 | 25 | 20 | 3.5 | 12.5 | 30% | 40% |
| 4/10 | 25 | 25 | 3.5 | 12.5 | 35% | 48% |
| 4/10 | 25 | 30 | 3.5 | 12.5 | 35% | 50% |

The peak efficiencies are extracted from the simulation results which are presented for the three different lenses in Figure S12-Figure S14. These results correspond to those presented in the main text Figure 3, but here with the full on-axis intensity maps shown for each lens separately with the color scale indicating the relative peak intensity as percentage of the incident power per µm$^2$. Integration of the intensity profile over the spot area yields the absolute focus efficiencies as given in panels C and D for each of the figures. Increasing the spot diameter (Area of Interest, AOI) from 8.8 to 52.8 µm gives an indication of how the intensity is distributed. The AOI sizes are given in multiples from 4x – 24x the simulation model step size of 2.2 µm (this is equal to the unit cell period).

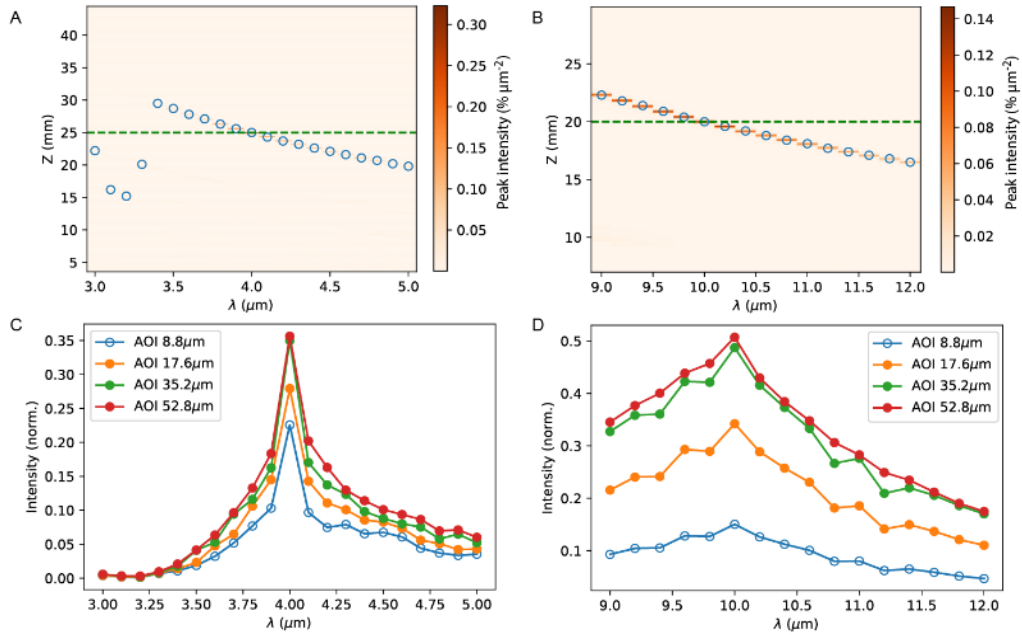

**Figure S12 A,B Simulated on-axis intensity profiles (color map) with maximum intensity (open circles) for MWIR and LWIR for the lens design with f4μm = 25 mm and f10μm = 20 mm. C,D peak efficiency (total intensity normalized to incident) within a circular area of interest (AOI) from 8.8 – 52.8 μm diameter, for MWIR and LWIR.**

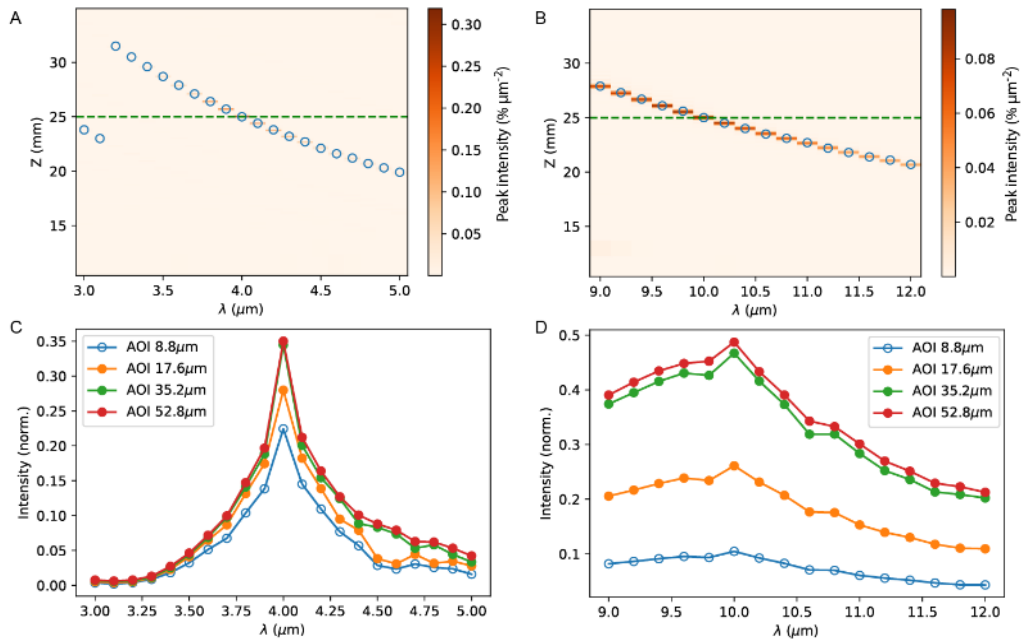

**Figure S13 A,B Simulated on-axis intensity profiles (color map) with maximum intensity (open circles) for MWIR and LWIR for the lens design with $f_{4\mu m}$ = 25 mm and $f_{10\mu m}$ = 25 mm. C,D peak efficiency (total intensity normalized to incident) within a circular area of interest (AOI) from 8.8 – 52.8 μm diameter, for MWIR and LWIR.**

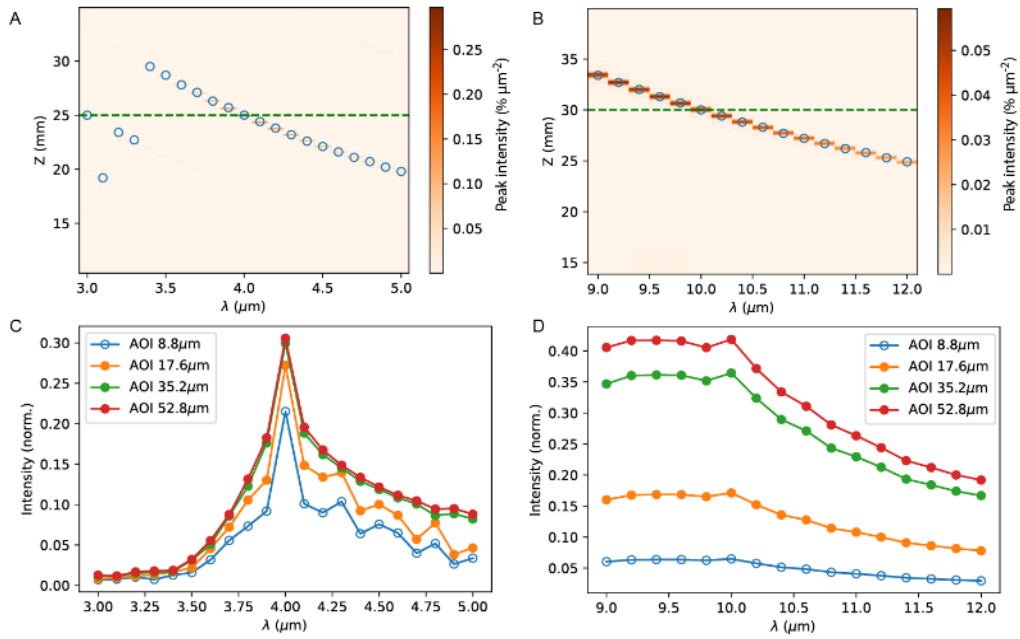

**Figure S14 A,B** Simulated on-axis intensity profiles (color map) with maximum intensity (open circles) for MWIR and LWIR for the lens design with f4µm = 25 mm and f10µm = 30 mm. **C,D** peak efficiency (total intensity normalized to incident intensity) within a circular area of interest (AOI) from 8.8 – 52.8 µm diameter, for MWIR and LWIR.

Details of the fabrication of metalens design V4 are given in Figure S15, showing the two DUV reticles of front and back side (A), photograph of the Nikon DUV lithography system and Track used for the fabrication (B), final produced wafer front and back side (C) and detailed photograph of back side (D).

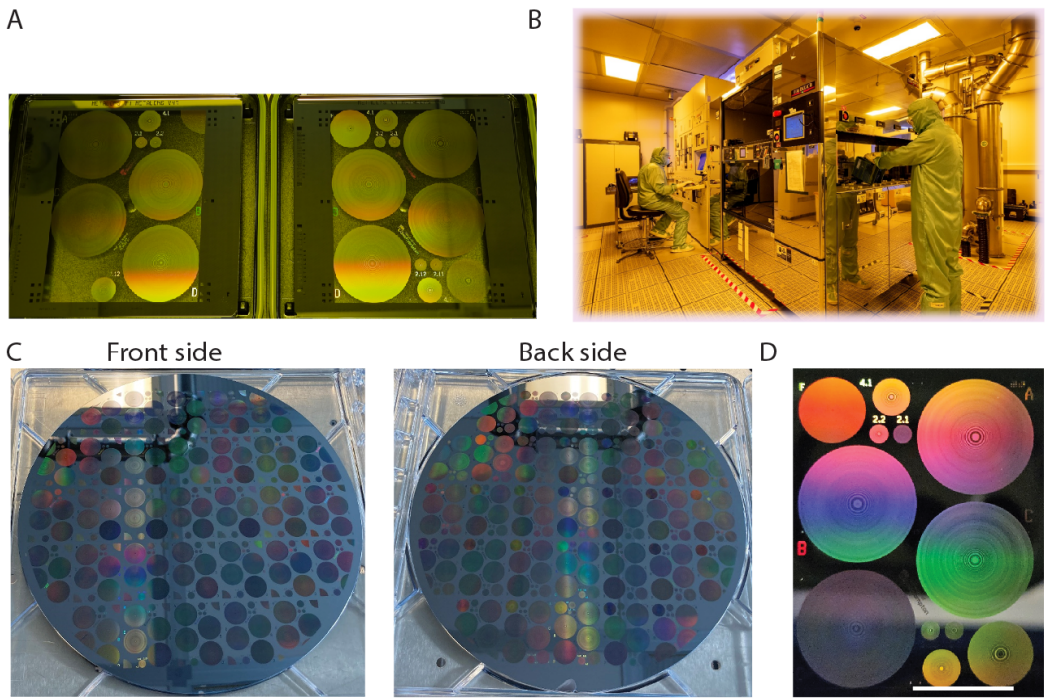

**Figure S15 A** pair of DUV reticles for front side and back side of metalens wafer V4. **B** Photograph of DUV Scanner and Track used for the double-side DUV lithography fabrication. **C** Fabricated wafer with 12.5 mm diameter double-sided meta-optics, showing front side and back side. **D** Zoomed-in optical photograph showing single write field with various metalenses, lenses labelled A, B, C are used in this study.

A full set of optical measurements results from wafer V4 is shown in Figure S16. The experimental setups are schematically shown in Figure S16A, for configurations where the 2.1×1.9 mm$^2$ MEMS blackbody light source is imaged directly using the metalens onto the mid-wave or long-wave camera (top / bottom setups), or where the blackbody source is used to illuminate a 1951 USAF test target (middle setup). We were unable to apply the USAF target imaging with the LWIR arrangement due to the presence of a SiO$_2$ substrate preventing transmission of blackbody in this region. The matching imaging lenses are Janos Nyctea 50mm for the MWIR camera and FLIR T197922 imaging lens for the LWIR camera. All images at different spectral band filters are presented in Figure S16B-D for the three types of metalenses. Figure S16C shows the detail of the blackbody light source where the bottom right corner was imaged on MWIR and the full area on LWIR, the difference in field of view being caused by the different imaging optics in the two setups.

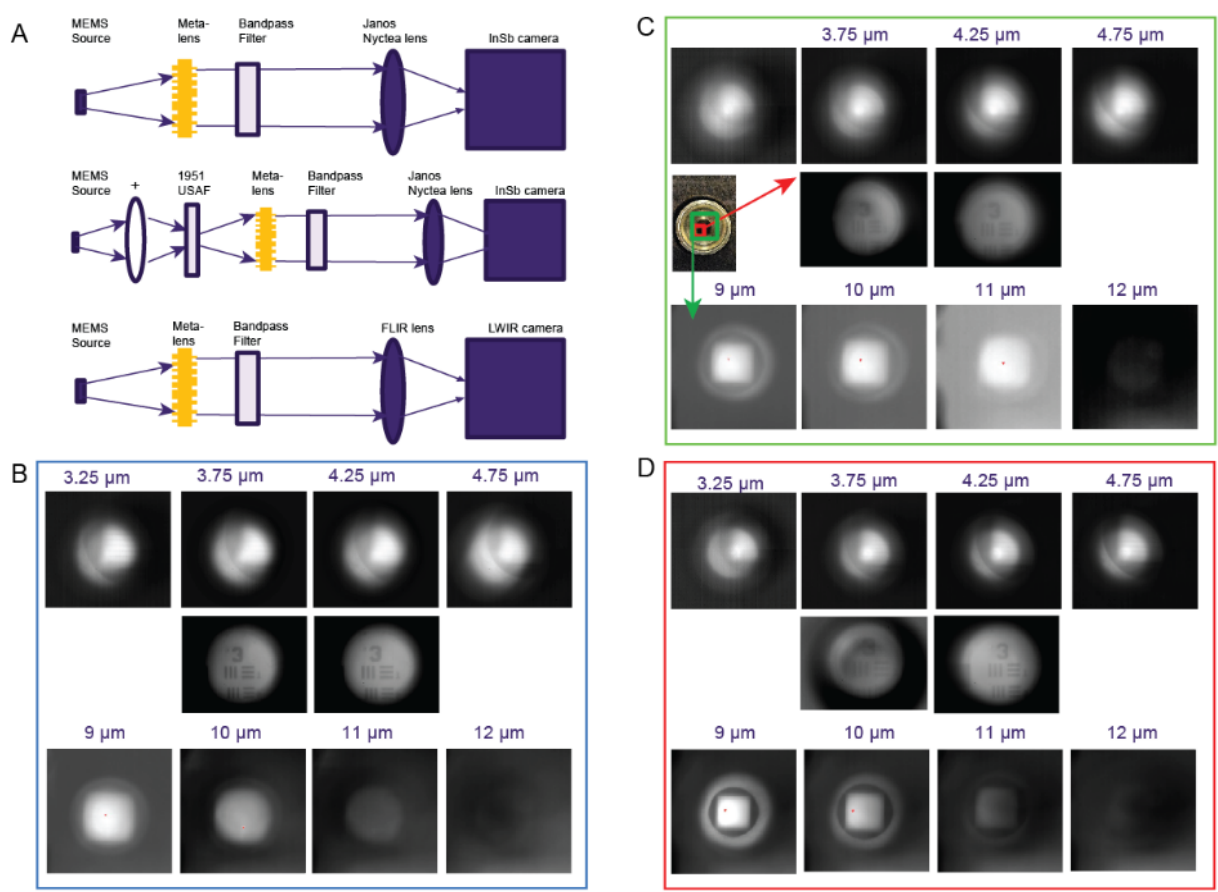

**Figure S16 A Schematic illustration of three different setups for measurement of the focal length using direct imaging of blackbody light source or of a 1951 USAF resolution target. B, C, D Full set of images obtained by direct imaging through a series of spectral bandpass filters in MWIR and LWIR range, for the three different metalenses of Table S2. Measurements of 1951 USAF target could only be obtained for wavelengths of 3.75 μm and 4.25 μm.**

Absolute focus efficiency (AFE) for the metalens V4 was obtained using a $CO_2$ laser at 10.8 µm wavelength. Figure S17A shows the experimental arrangement, where a 200 µm pinhole was used to filter the focused light from unfocused background. The focus was imaged onto the LWIR camera using a set of neutral density filters to strongly attenuate beam in order to prevent damage to the uncooled sensor. Focal spots are shown in Figure S17B for the three different metalenses showing a focus of around 2-3 pixels on the camera. Using the same arrangement, the pinhole was placed in the focal plane and aligned to transmit the laser spot. After alignment, an optical power meter was placed behind the pinhole to measure the transmitted power. This power was normalized to the incident power without pinhole and metalens to obtain the AFE as shown in Figure S17C. AFE values of around 25% were obtained experimentally, compared to simulated values of between 30%-35% at the wavelength, indicating a performance of the metalens which is in reasonable agreement with the design.

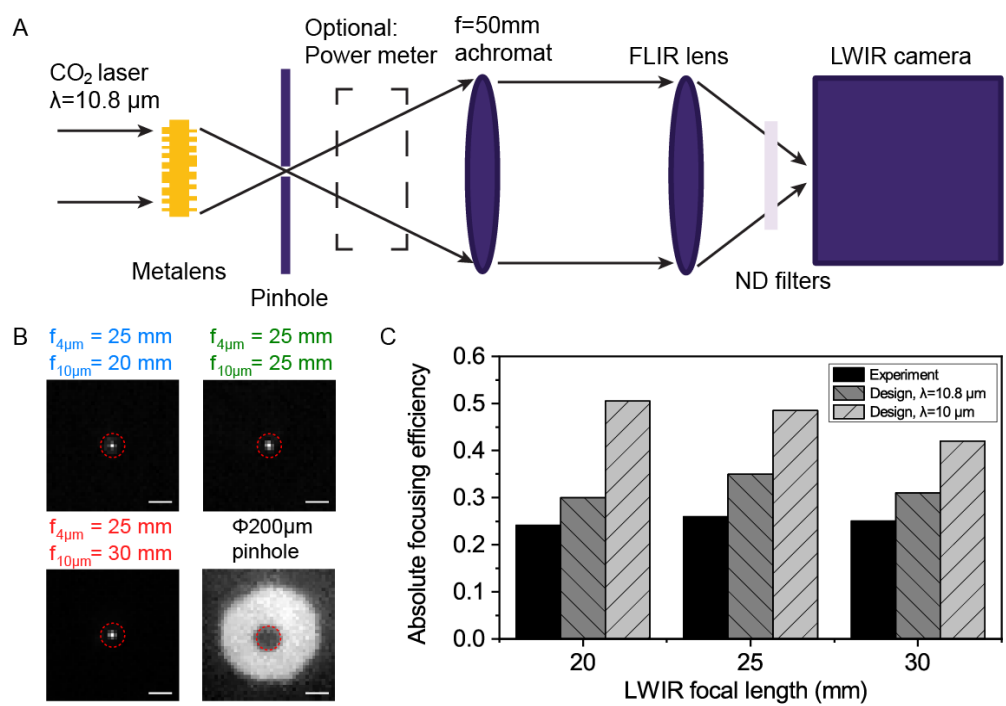

**Figure S17 A Schematic of setup for measuring of absolute focusing efficiency using $CO_2$ laser at 10.8 µm wavelength. B Images obtained using LWIR camera of focus spot produced by metalens using achromatic collection optics, as well as of pinhole surface placed in focal plane of collection lens. Scale bars are 200 µm. C Absolute focusing efficiency experimentally determined using power meter placed after pinhole, compared with simulated design performance at 10.8 µm and 10 µm wavelength (maximum of design).**

## S8. Imaging of University of Southampton logo test sample

Details of the University logo test sample are shown in Figure S18. The logo test sample is designed as a three-layer stack (Figure S18A) using a Transformation Matrix method with the peak emissivity (absorption) at 4 µm. The stack consists of 80 nm Al:ZnO, 1000 nm $SiO_2$ and 100 nm Al as back reflector. The Al:ZnO (AZO) was grown using atomic layer deposition with TMA, DEZ and DI water as precursor and a TMA:DEZ cycle ratio of 1:24, at a temperature of 250 °C. The $SiO_2$ and Al were grown using a Helios sputtering system. Detailed process information is available in our previous publication, Sun *et al.* Adv. Mater. 2020, 2001534. The infrared response is measured by a FTIR system, resulting in the emissivity spectra presented in Figure S18C with the measurement locations detailed in Figure S18B.

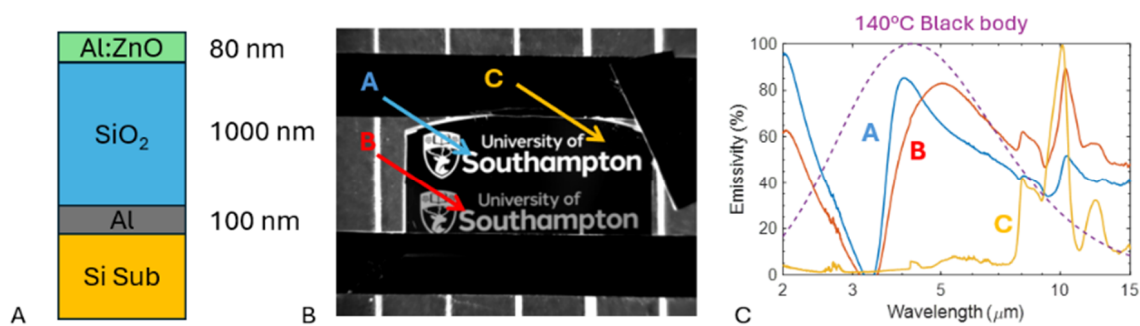

**Figure S18 University logo structures and infrared response. (a) three-layer stack with high conductive Al:ZnO on $SiO_2$ on Al, (b) Thermal imaging with locations labelled, (c) Emissivity spectra measured at labelled locations, with black body spectra presented at 140 °C.**

Figure S19 shows the experimental setup for imaging of a heated test object, for the long-wave (A) and mid-wave (B) range. The infrared camera is positioned at approximately 42 cm distance from the test sample. For the long-wave setup, the hybrid optical stack consists of a 25 mm diameter $BaF_2$ refractive lens and 40 mm diameter metalens which are independently mounted. The $BaF_2$ optic is retained in a 2-inch diameter lens tube and the metalens in a 42mm diameter T-mount lens tube. The two tubes slide into each other to provide a compact tunable optical system with relative positions controlled by using two manual translation stages. The $BaF_2$ optic is positioned very close to the LWIR camera at an estimated distance of around 15mm from the focal plane array, this was possible due to the uncooled camera offering good access close to the focal plane array. For the mid-wave setup, the cooled InSb camera CEDIP/FLIR SC7300 Titanium has a cold chamber with limited access, therefore the hybrid refractive and metasurface configuration could not be used. In this case, the metalens was used by itself in a singlet configuration.

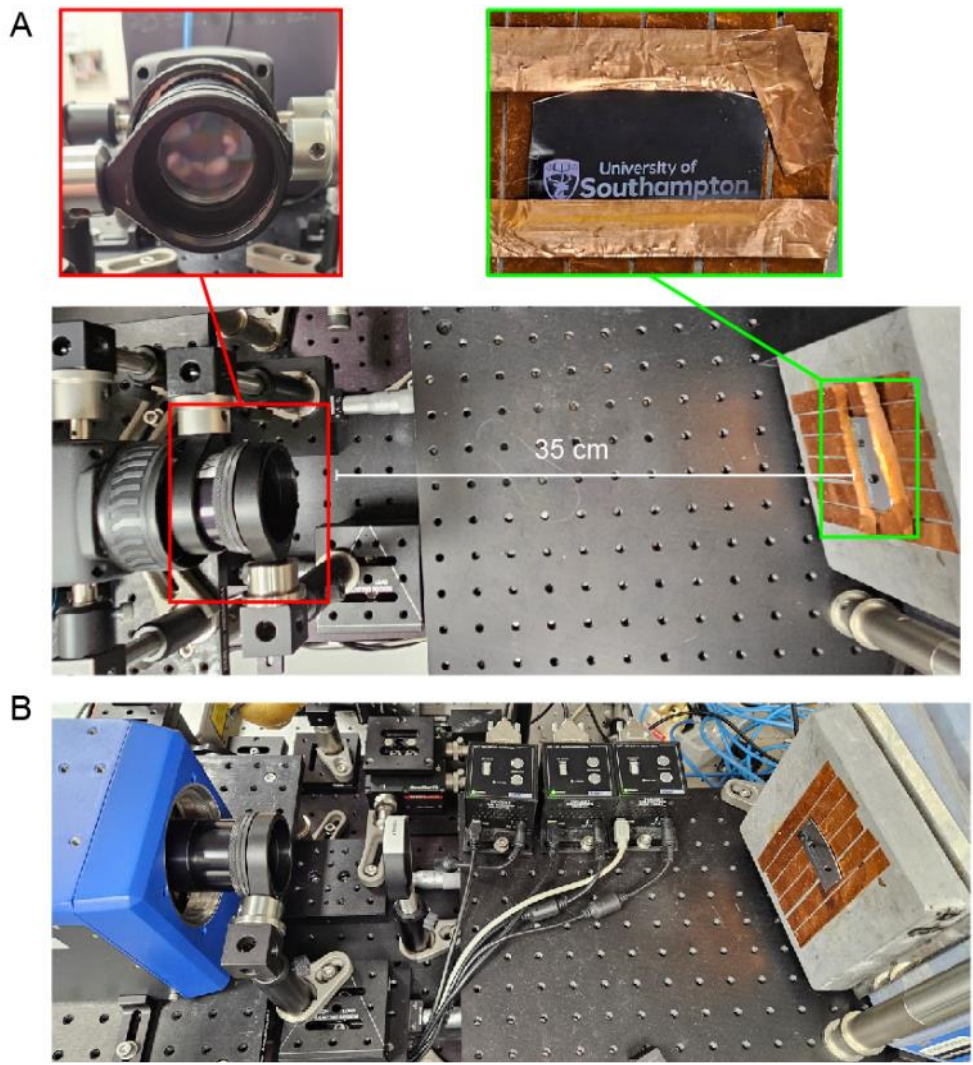

**Figure S19 Experimental setup for imaging of Logo test target at 140 °C. (a) Long-wave imaging setup using FLIR A655sc LWIR camera, hybrid optical system consisting of BaF$_2$ refractive lens and metalens (front of metalens shown in inset) and vertically mounted hot plate with sample attached using copper tape. (b) Setup for Mid-wave imaging using CEDIP/FLIR Titanium camera with metalens (no BaF$_2$ refractive lens).**

**S9. Hybrid meta-optic and refractive compound lens**

To put into perspective the performance of the hybrid metalens – refractive BaF$_2$ configuration, we performed the same LWIR imaging of the target object using just the metalens without refractive optic. Figure S20A, B show results for both the dual-band (A) and LWIR-only (B) metalenses. For both metalenses, the images are unsharp and with very low contrast on the university logo. The outlines of the copper tapes are also much less clear and the lines of emission between the adjacent copper strips are hazy. We attribute this difference to the lack of chromatic correction for the metalenses over the bandwidth, thus showing the good performance in comparison by the hybrid metalens – refractive optical system of which results are presented in the main text.

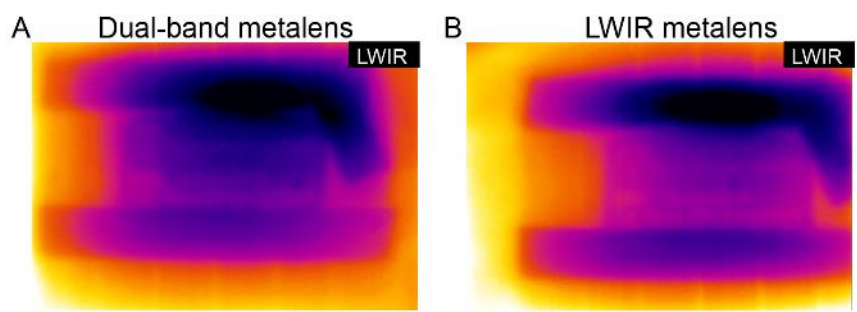

**Figure S20 Images of target object heated to 140°C for singlet metalenses without the BaF$_2$ hybrid arrangement, for dual-band metalens (V5) and LWIR metalens (V6).**

**S10. Additional imaging of dual-band metalens**

We have tested the fabricated 40 mm dual-band metalens V5 in real-life imaging with MWIR/LWIR cameras. MWIR imaging was performed for a heated soldering iron at 150°C with the black-body emission in the MWIR band and presented in Figure S21a comparing with the camera's original Nyctea imaging lens Figure S21b. The Metalens imaging shows clearly the soldering iron details. Compared with the original Nyctea lens imaging, the metalens image is a bit blurry. This is attributed to the chromatic aberration of metalens.

LWIR imaging was performed for a human hand with images shown in Figure S22 for the 40 mm dual-band metalens (a), commercial CaF$_2$ lens (b) and commercial BaF$_2$ lens (c), both refractive lenses are 25mm in diameter with 50mm focal length and f-number F/2. All three singlet lenses have strong chromatic aberrations individually, with the CaF$_2$ having a higher dispersion than the BaF$_2$ lens. To achieve improved chromatic aberration control, we adopted the developed compound system with one metalens and one CaF$_2$ or BaF$_2$ lens for aberration correction, with resulting images are presented in Figure S22. The metalens/CaF$_2$ hybrid (Figure S22a) gives a good quality image in LWIR band with an approximately 2 times larger field of view than the individual F/2 singlets. A similar improvement is also seen using the Metalens/BaF$_2$ hybrid combination (Figure S22b) which achieves a higher signal to noise due to the higher overall transmission of the BaF$_2$ lens. Figure S22c shows a LWIR image of the soldering iron at 150°C using the Metalens/BaF$_2$ hybrid. These real-world imaging demonstrates that the compound solution can offer an effective way to correct metalens aberration and provide good imaging.

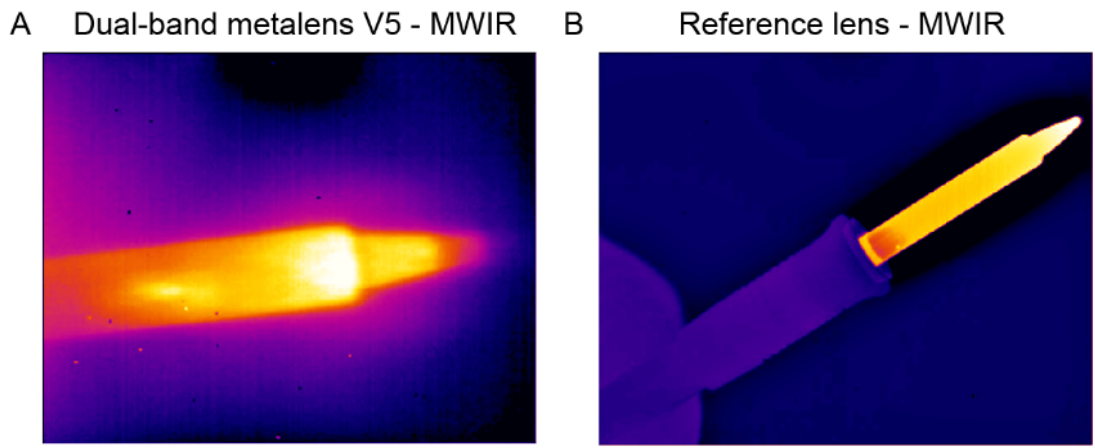

**Figure S21 MWIR imaging measurement through InSb camera for a 150 ℃ soldering iron at a distance of 1 m from the camera, by dual-band metalens V5 (a) and the Nyctea 50 mm lens (b).**

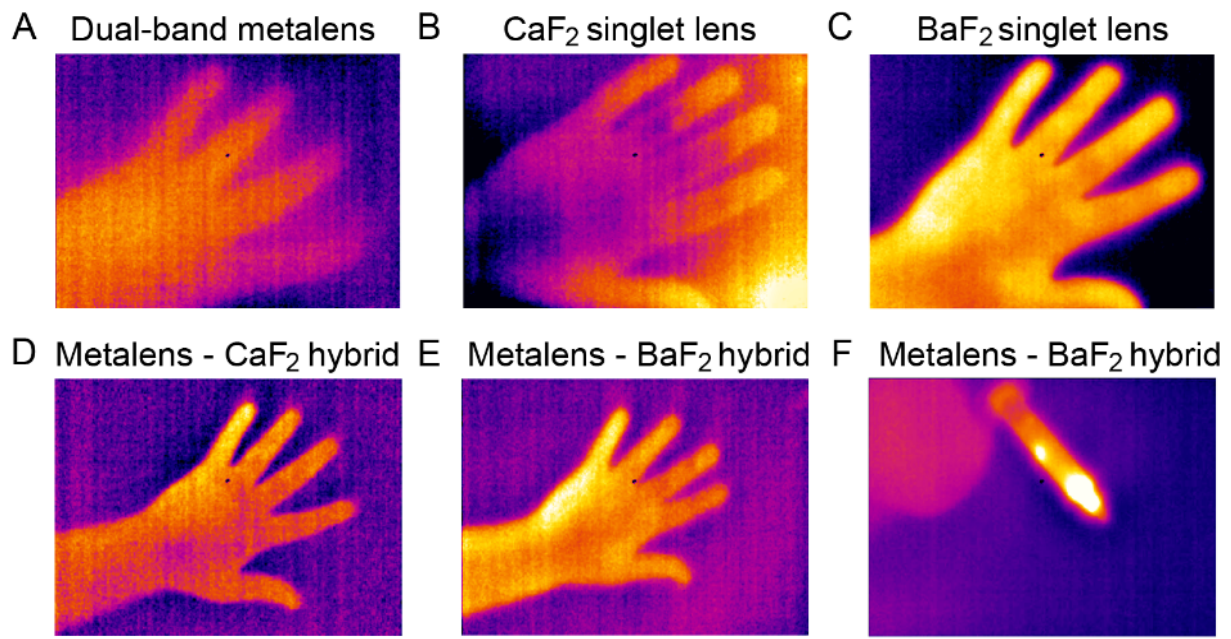

**Figure S22 (a-c) LWIR imaging measurement using singlet lenses for dual band metalens (a), CaF$_2$ lens (b) and BaF$_2$ lens (b). (d-f) Compound system imaging using metalens - CaF$_2$ hybrid (d), metalens-BaF$_2$ hybrid. Hand in (a-e) was kept in same position and distance for all images. (f) LWIR image of soldering iron at 150 ℃ using metalens-BaF$_2$ hybrid.**

## S11. Outdoor LWIR demonstrator

For the outdoor demonstrator the LWIR camera system was set up pointing towards a scene with objects at distances from 10 – 150 m as illustrated in Figure S23A, B. Figure S23A illustrates the local topography of the Highfield campus at the University of Southampton, where the camera is located in the corner of building 46 and pointing in the direction of building 40. The roof of building 40, located at around 150 m distance, is clearly seen in the optical photograph of Figure S23B showing the elevation of this structure. As the optical image is taken at a slightly higher vantage point than the infared demonstration, the alignment of the person with the roof is not present here, as this is seen more clearly in the infrared images which are taken from the height of an optical bench. The infrared images are optimized to maximize the visibility of the branches of the trees, located at around 25 m distance. Figure S23C-F shows the raw and processed images from the metalens – refractive BaF$_2$ hybrid optical system.

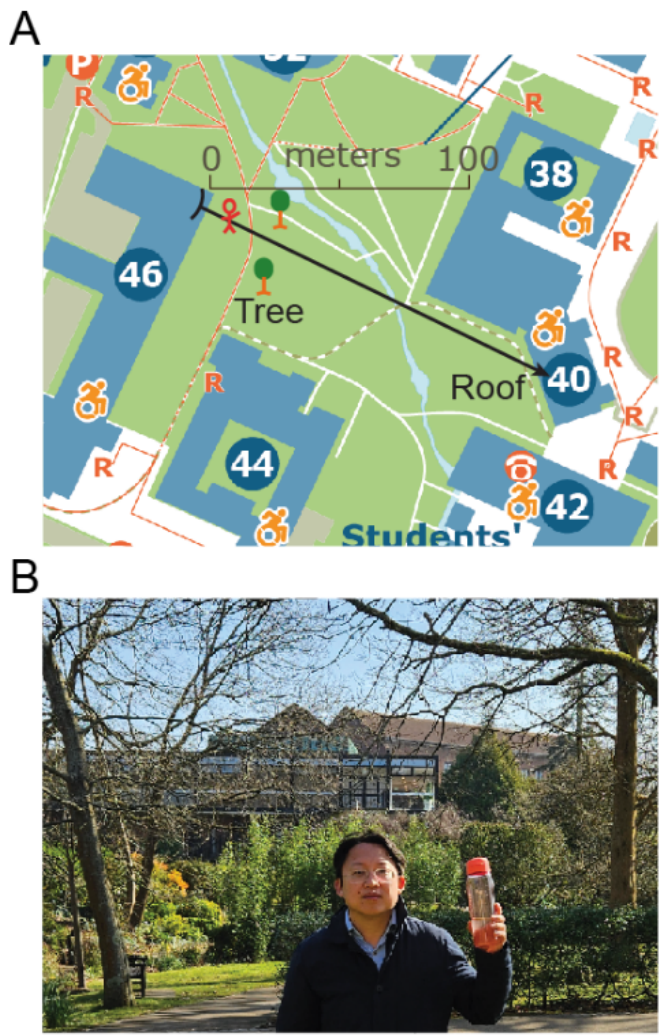

**Figure S23 A Map detail showing geographic arrangement to scale with camera position, person at 10m, trees at 25 m, and rooftop at 150 m distance. B Visible-band photograph of scene used for outdoor demonstrator.**

## S13. Performance testing of large-area metalenses

Lens performance was evaluated by Teledyne Qioptiq Ltd using specialized setups commonly used in the optical system manufacturing industry.

LWIR analysis.

RMS. RMS quantifies the overall wavefront error, with $0.042\lambda$ indicating relatively smooth deviations compared to the wavelength. A low RMS value implies high optical quality, aligning with the Strehl ratio of 0.93 (see below). Interpretation: The optical system has minor residual imperfections.

Peak-Valley. PV measures the extreme deviations in the wavefront error. For the experimental data, $0.28\lambda$ shows slightly higher error than the simulation ($0.2467\lambda$). Experimental vs. Simulation: The slight difference suggests environmental factors, manufacturing errors, or alignment inaccuracies influencing the experimental data.

Strehl Ratio. A Strehl ratio of 0.93 indicates excellent optical performance (ideal Strehl ratio = 1 indicating that an optic is perfect and aberration free). At SR>0.8, the system is diffraction-limited, meaning its performance is governed by fundamental physics rather than imperfections. Interpretation: The wavefront quality allows for near-ideal focusing.

Astigmatism. This represents cylindrical distortions, with a magnitude of $0.06\lambda$, oriented at -87.1°. This small value implies minimal elongation of the point spread function (PSF). Interpretation: Astigmatism is a minor contributor to the overall error.

Coma. Coma introduces asymmetric distortions in the PSF, typically caused by off-axis angles. A value of $0.27\lambda$ at 153° indicates moderate coma contribution.

Spherical. Spherical aberration is the largest contributor among measured aberrations, at $0.30\lambda$. This results from the inherent curvature of optical surfaces, causing edge rays to focus differently from central rays. Interpretation: Spherical aberration limits the Strehl ratio slightly.

Experimental vs Simulation. The experimental PV ($0.28\lambda$) is close to the simulation PV ($0.2467\lambda$), indicating the system behaves predictably under real conditions. Discrepancy: Slight deviations may stem from environmental factors, material inhomogeneities, or manufacturing tolerances.

Summary. The low RMS (0.042) and high Strehl ratio (0.93) confirm excellent wavefront quality. Residual errors are dominated by spherical aberration (0.30), followed by coma (0.27), with minor contributions from astigmatism (0.06). Experimental results closely match simulation, validating the model's accuracy.

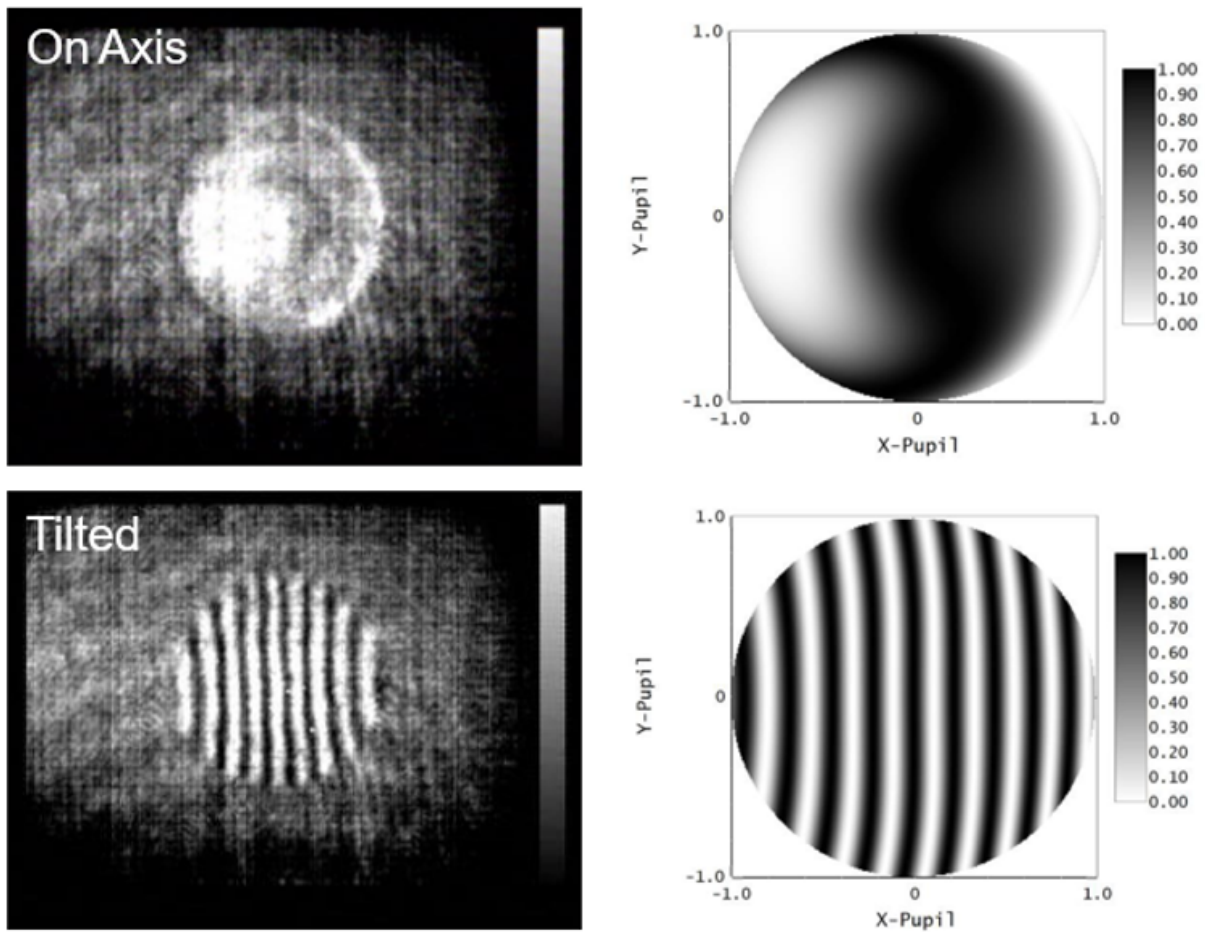

**Figure S24 LWIR interferometry results at 9.24 μm wavelength, showing on-axis defocus and off-axis tilt, experimental (left) and simulations (right) for a peak-valley of 0.2467λ.**

MWIR analysis.

The wavelength of 3.39 μm was dictated by the availability of a light source at this wavelength which is quite far from the design wavelength of the metalens. The measurements were done by removing / adding power at focus. Removing power at focus isolates higher-order aberrations (like astigmatism, coma, and spherical aberration) by eliminating the dominant curvature effects. This allows the analysis to focus on the residual wavefront errors without contributions from lens focusing power.

Purpose: Removing power at focus simplifies the evaluation of optical imperfections, as curvature can dominate the wavefront error and mask subtler aberrations.

Implication: This is particularly important in metrology setups where higher-order aberrations, not corrected by removing power, are the focus of analysis.

Deliberately introducing defocus, especially with power, allows for controlled testing of optical behaviour under non-ideal conditions, which can simulate real-world scenarios or stress-test optical performance.

Purpose: To evaluate the robustness of the system's aberrations, Strehl ratio, and RMS wavefront error when subjected to misalignment or off-axis effects.

Implication: This configuration exaggerates contributions from tilt, astigmatism, coma, and spherical aberration, providing insights into how the lens performs when improperly aligned.

RMS. The RMS wavefront error quantifies the standard deviation of the optical wavefront deviations from an ideal flat wavefront. A value of $0.051\lambda$ is measured, which indicates minimal deviation, suggesting high optical quality. Significance: For high-performance systems, an RMS error below $0.07\lambda$ is typically considered diffraction-limited, where system performance is primarily governed by the diffraction of light rather than aberrations or imperfections.

Peak-Valley. $PV = 0.239\lambda$: The peak-valley metric represents the difference between the maximum and minimum wavefront deviations. A low PV value like this suggests minimal surface or optical imperfections. Limitations: While PV provides an overall sense of optical error, it is less robust than RMS because it can be dominated by localized defects.

Strehl Ratio. A Strehl ratio of 0.9024 confirming optical system is diffraction limited as it exceeds the threshold of 0.80.

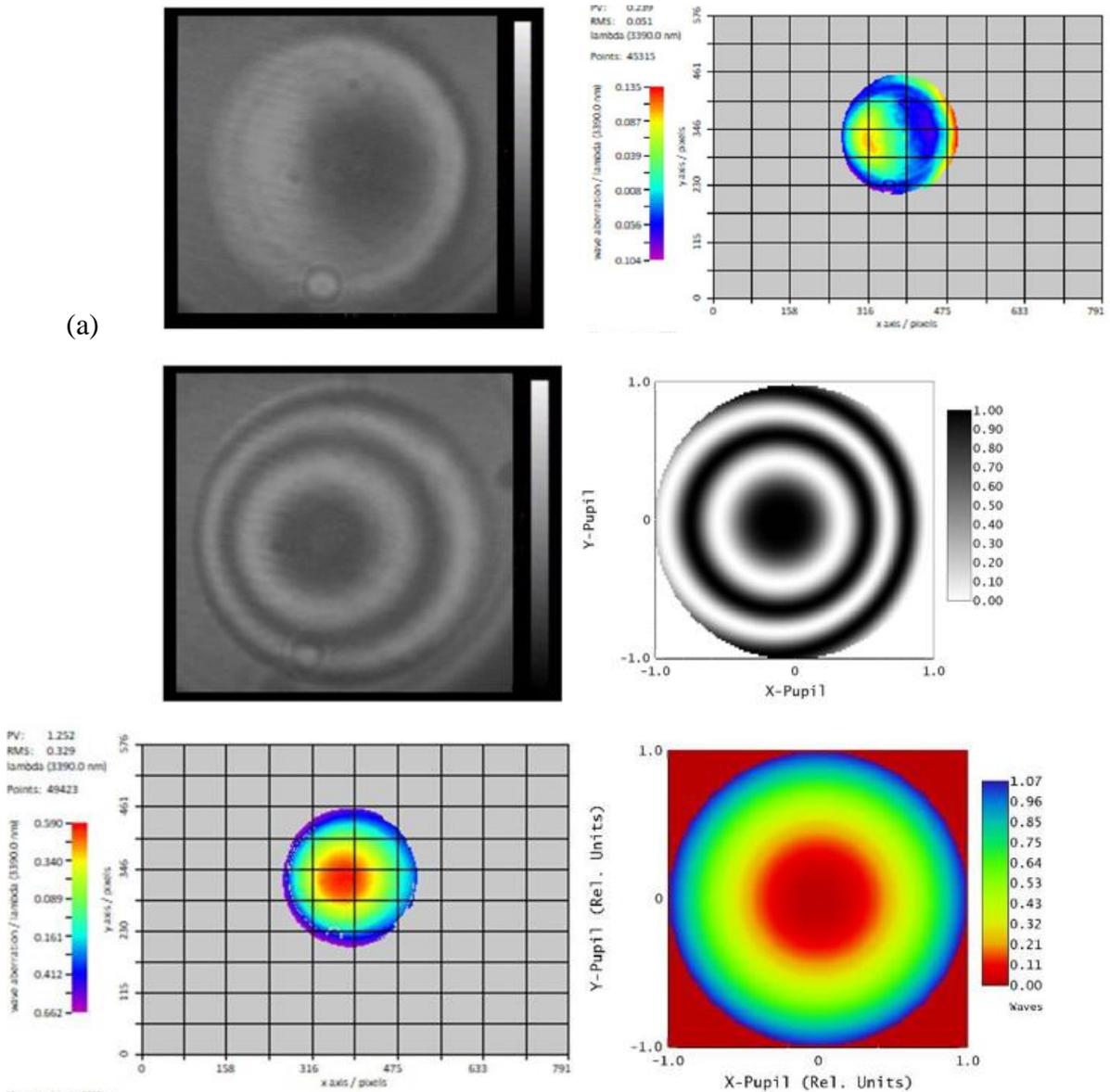

(a)

**Figure S25 MWIR interferometry results at 3.39 μm wavelength, showing on-axis defocus and off-axis tilt, experimental (left) and simulations (right) for a peak-valley of 0.2467λ.**

Astigmatism. Astigmatism = 0.56λ (at -18.1°). This is a second-order aberration where the wavefront has different curvatures along perpendicular axes, causing lines in one orientation to focus differently than lines in another. Astigmatism of this magnitude indicates moderate asymmetry, likely due to lens imperfections or stress during mounting, or the purposely added tilt.

Coma. Coma = 1.02λ (at 5.9°). Coma is a third-order aberration caused by off-axis light rays forming comet-shaped blur spots. The relatively high value of coma indicates significant asymmetry, due to misalignment with the purposely added tilt.

Spherical. Spherical = 0.06.: Spherical aberration arises when marginal rays (closer to the edge of the lens) and paraxial rays (near the center) focus at different points. A value of 0.06 indicates a negligible contribution, suggesting high-quality lens design.

Summary. The low RMS and PV values, combined with high Strehl ratio, indicates the lens has good optical quality and is close to diffraction-limited performance.

Coma and astigmatism dominate the residual wavefront error, likely due to the purposely introduced tilt causing alignment issues.

Spherical aberration is negligible, suggesting the lens is well-corrected for this high-order aberration.

## S12. Removal of bad pixels and defects

In the main text, some minor defects in the images have been removed by replacing them with adjacent colors, as illustrated in Figure S26. No physical information is contained in these defects and all corrections made are cosmetic in nature. The cluster of around 10 dark pixels is an artefact from $CO_2$ laser damage on the camera produced by an overexposure incident during the focus measurements of Figure S17. Small arrows in the outdoor imaging test automatically generated in the imaging sofware were manually removed in the same way.

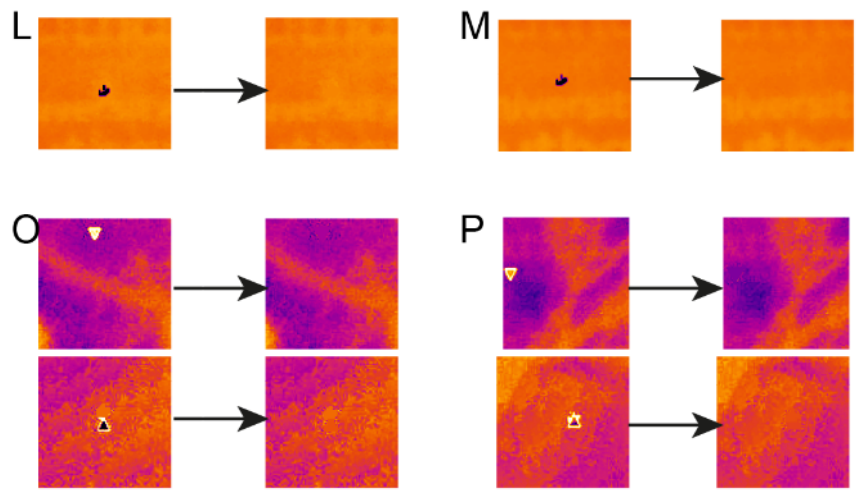

**Figure S26 Details of images in main text Figure 4L, M, O, P before and after removal of bad pixels and markers.**